# A novel method of experimental determination of grain stresses and critical resolved shear stresses for slip and twin systems in a magnesium alloy

P. Kot[1], M. Wroński[2,*], A. Baczmański[2], A. Ludwik[2], S. Wroński[2], K. Wierzbanowski[2], Ch. Scheffzük[3], J. Pilch[4] and G. Farkas[4]

[1]NOMATEN Centre of Excellence, National Centre of Nuclear Research, A. Sołtana 7, 05-400 Otwock-Świerk, Poland

[2]AGH University of Krakow, Faculty of Physics and Applied Computer Science, al. Mickiewicza 30, 30-059 Krakow, Poland

[3] Karlsruhe Institute of Technology, Institute of Applied Geosciences, Adenauerring 20b, 76131 Karlsruhe, Germany

[4]The Czech Academy of Science, Nuclear Physics Institute, Hlavní 130, 250 68 Řež, Czech Republic

(*) Corresponding author

## Abstract

A novel original method of determination of stresses and critical resolved shear stresses (CRSSs) using neutron diffraction was proposed. In this method, based on the crystallite group method, the lattice strains were measured in different directions and using different reflections *hkl* during uniaxial deformation of magnesium alloy AZ31. The advantage of this method is that the stresses for groups of grains having similar orientations can be determined directly from measurement without any models used for data interpretation. The obtained results are unambiguous and do not depend on the models' assumptions as in previous works. Moreover, it was possible for the first time to determine the uncertainty of the measured CRSS values and local stresses at groups of grains.

The used methodology allowed for the determination of stress partitioning between grains having different orientations and for an explanation of the anisotropic mechanical behaviour of the strongly textured alloy. Finally, the CRSS values allowed for the validation of the type of intergranular interaction assumed in the elastic-plastic self-consistent model and for a significant reduction of the number of unknown parameters when the model is adjusted to the experimental data.

## 1. Introduction

Magnesium alloys, due to their unique properties such as high specific strength, high thermal conductivity and damping capacity, have been used in many industries. However, the limited ductility and poor formability of Mg alloys at room temperature restrict their wide range of applications [1,2]. The poor formability is a consequence of the strong plastic anisotropy of Mg alloys. In the hcp phase of Mg alloys, plastic deformation occurs by means of the following crystallographic slips: basal <a> $\{0001\}\langle 11\bar{2}0\rangle$, prismatic <a> $\{1\bar{1}00\}\langle 11\bar{2}0\rangle$ and the pyramidal slips: <a> $\{1\bar{1}01\}\langle 11\bar{2}0\rangle$ , first order <c+a> $\{1\bar{1}01\}\langle \bar{1}2\bar{1}3\rangle$ and second order <c+a> $\{1\bar{2}12\}\langle \bar{1}2\bar{1}3\rangle$ which have very different values of critical activation stresses. The listed slip modes are generally not sufficient to accommodate an imposed plastic deformation, therefore in Mg alloys the activation of twin modes is also observed, realized by the tensile twin mode $\{1\bar{1}02\}$ and compression twin mode $\{1\bar{1}01\}$. Various modifications of the chemical composition and microstructure can be used to weaken the strong plastic anisotropy, which in consequence improve the formability of Mg alloys, such as:

(i) alloying - the addition of rare-earth elements (e.g. Y, Ce, Nd, La) [3–5] weakens the crystallographic texture of Mg alloys and enhances the activation of non-basal slips,

(ii) texture modification via different processing - e.g. [6] has shown that the asymmetric rolling process improves the formability of Mg alloys by suppressing the twinning activity,

(iii) and grain refinement - reducing the grain size suppresses the twinning activity, leads to a homogeneous microstructure, and improves formability [7–9].

All these modifications can significantly affect the deformation behaviour of the newly obtained magnesium alloy. Therefore, it is important to develop test methods to study changes in material properties caused by such modifications. It is particularly important to determine the critical resolved shear stresses (CRSSs) necessary for activation of slip and twin systems, as well as to measure the stresses at polycrystalline grains and to determine the plastic deformation of the grains. It is obvious that knowledge of the CRSS and plastic processes occurring at the scale of polycrystalline grains is necessary to understand and describe material properties such as hardness, hardening and elastic-plastic behaviour at the macroscopic scale.

The macromechanical properties of polycrystalline materials were determined by means of mechanical tests and an effort was made to estimate the measurement uncertainty of such values as the stiffness modulus [10], compressive [11] or flexure strength [12], as well as residual stresses [13–15]. Characterization of properties at the level of polycrystalline grains, however, is usually more complex, and the uncertainty of values such as the CRSS is difficult to estimate. The CRSS values estimated for the AZ31 alloy exhibiting different texture and microstructure are often scattered in wide ranges, which, after eliminating extreme values, are as follows: for basal slip systems 2-50 MPa, for prismatic systems 55-100 MPa, for pyramidal systems 50-100 MPa, and for tensile twins 15-50 MPa (Suppl. Material in [16]). It should be emphasised that the values are given without uncertainties. The value of 200 MPa for compression twins can be added to this review [17]. Generally, it can be stated that the CRSS of the deformation modes of magnesium alloys can be ordered from low to high values as follows: basal slip, tensile twinning, prismatic slip, pyramidal slip, and compression twinning.

Accordingly, two easy deformation modes are basal slip and tensile twinning. However, the basal slip, having the lowest CRSS, is predominant only in advantageous crystal orientations, when loading force is deviated from the crystal <c> axis and from the basal plane. Therefore, other slip or twin systems must also be activated to accommodate plastic deformation [18,19]. Moreover, twinning can be initiated by the processes occurring in the neighbouring grains (autocatalytic twin nucleation [20]) leading to the appearance of shear bands [20]. It was also found that so-called slip-induced tensile twinning appears, which is mainly governed by the geometrical relation between the source slip system and the activated twin system [21]. Another interesting result is presented in [22], where the visco-plastic model was developed by including drag stress for dislocations and dislocation density hardening and material was treated as a composite of grains and twins. It was found that drag stress for prismatic slip is very sensitive to temperature in contrast to basal and pyramidal slip. Different factors can influence the deformation pattern in magnesium and magnesium alloys. Besides temperature, these also include ageing [23]. It was found that in the initial range of deformation, ageing promotes twinning over slip, but the presence of precipitates leads to thinner twins. Upon unloading, further forward twinning as well as de-twinning are observed.

The CRSS values can be estimated on the basis of hardness measurement [24,25], e.g. in paper [26] the nanoindentation technique was used to study the influence of the alloying element Nd on the CRSS values of various sliding systems in the MN11 alloy. The nanoindentation technique has been intensively developed in the last decade and recently an attempt was made to analyse the problem of CRSS uncertainty [25]. The main advantage of this technique is that the elastic-plastic properties can be tested using a very small test volume of the material [27,28]. Moreover, the technique is very fast compared to conventional macroscopic examination techniques such as compression or tensile testing. However, the hardness measurement only provides information about mechanical properties at the surface and a shallow depth below the surface. In addition, the use of crystallographic models is necessary to interpret the experimental results [26]. In general, this method is used to test the mechanical properties and residual stresses in the near-surface zone modified by, among others, shot-penning, surface mechanical attrition treatment (SMAT) or layer deposition [29,30].

Another approach to the study of deformation mechanisms in magnesium was proposed using simulations based on the molecular dynamics (MD) method [31,32]. The calculations developed using MD are directly comparable to experimental results of mechanical tests on very small samples. It should be noted that a series of interesting observations were obtained by performing compression tests on micro- and nanopillar magnesium samples [33–35]. In the case of compression along the <c> axis of micropillar samples, pyramidal slip was observed and subsequently, due to a misalignment of loading force, a massive basal slip was observed. Additionally, compression twinning was observed in nanopillar samples, followed by basal slip inside a formed twin. On the other hand, when tensile strain was imposed along the <c> axis, tensile twinning was activated [34]. Next, when the crystal <c> axis was strongly tilted from the loading direction, the basal slip was predominant. And finally, tensile twinning dominated when compressive stress was applied perpendicularly to the <c> axis [35]. The calculations made using MD generally confirm the above experimental results.

Acoustic emission is another useful technique for studying the micromechanical behaviour of materials, especially the twinning nucleation process [36–41]. Acoustic emission occurs during sample deformation and may come from various sources that are difficult to distinguish. One of the first and most popular methods based on acoustic emission is the so-called hit-based method, in which the full duration of the examined event can be recorded. A characteristic feature of acoustic emission is the registration of sound waves coming from different sources, which allows these sources to be identified using spectral analysis. An example of a technique that uses this dependency is called adaptive sequential k-means (ASK) [41,42]. This method was used to test a magnesium alloy and allowed to demonstrate significant activity of the primary and twin systems during mechanical tests carried out in different directions for a sample with a significant texture [38,39]. It is worth noting that acoustic emission is a method complementary to diffraction studies, especially neutron diffraction (presented below). The combination of these methods allows the mechanisms of plastic deformation to be described and the dynamics of these processes to be studied [41].

The above discussion shows that, in the case of the nanoindentation method used to determine CRSS, the results are obtained for the near-surface volume, while the results of the acoustic method are rather qualitative. Therefore, the commonly used method for determination of the CRSS value is based on diffraction measurements of the lattice strains during mechanical tests, such as tensile or compression tests. In principle, the CRSSs can be determined directly from the experiment if the stress state in the polycrystalline material is measured at the grain scale. Various methods have been used to measure residual stresses, i.e. stresses remaining in the material after processing [13,15,30,43,44]. The most direct methods for determining the residual and applied stresses in polycrystalline materials are based on the measurement of lattice strains using neutron or X-ray diffraction [13,15,43,45–49]. From the point of view of the representativity of results, measurements for a large volume containing a large number of polycrystalline grains are particularly useful. Such measurements are possible using neutron radiation [13,50,51] or high-energy synchrotron radiation [13,49,51], which additionally enable measurements inside the sample. One of the very important advantages of the diffraction technique is the ability to measure stresses for individual phases of polycrystalline materials [52–54] and even for individual grains or groups of grains [13,55–59].

An interesting methodology for determination of the grain stresses and then CRSS values was presented in [59], where the lattice strains in Ti-8.6Al alloy were measured using monochromatic high energy synchrotron radiation (diffraction microscopy technique) during a tensile test. The measurement was performed in transmission mode and the sample was rotated around the loading direction. The evolution of stress tensor was determined from the lattice strain measured for 421 large grains having an average size of 106.8 μm. Then the dependence of the maximal resolved stresses (RSS) on four slip systems vs. applied macroscopic stress was presented. It was assumed that the CRSS values can be determined from such a relationship, averaged for all grains, as the point at which the trend of the graph changes when the slipping system is activated. However, it should be emphasised that the discrepancy of the resolved shear stresses for different grains is large, and the uncertainty of the determined values is not given. In addition, this methodology cannot be applied for small grains of a dozen μm or less, which are typically found in polycrystalline materials subjected to mechanical processing such as rolling.

In [55], the resolved shear stresses for slip systems activated in pure titanium were determined directly from the experiment using a similar method as above (diffraction of high energy synchrotron radiation). The measurements were carried out for a limited number of single grains with an average size of about 100 μm. The CRSS values for basal and prismatic slip systems were determined based on the evolution of the RRS and grain orientations as a function of the applied load. To do this, from all available grains, those that exhibited lattice rotation either about the <c>-axis or an axis perpendicular to it (being indicative of exclusive prismatic or basal slip, respectively) were checked for their maximum RSS on the presumed slip systems. Using this method, CRSS values were determined for only two slip systems, and the results are not representative of a large number of grains. Such methodology cannot be used for a sample containing fine grains, which is the case of plastically deformed polycrystalline materials.

The method usually used to determine CRSS values and hardening parameters for large statistics of small grains is the TOF (time of flight) neutron diffraction experiment, e.g. [17,19,54,60–66]. During the tensile/compression test, the lattice strains are measured *in situ* in two directions of the scattering vector, i.e., along the applied load and in the transverse direction. For each direction, the measurement is performed using many reflections *hkl*, i.e., the average strains of the lattice are measured for grain volumes for which the scattering vector is perpendicular to the (*hkl)* plane. Unfortunately, on the basis of the data obtained in this way, stresses in individual grains or groups of grains cannot be directly determined from experiment and an elastic-plastic model of a polycrystalline material must be used to find the variables characterizing crystallographic slips and the twinning process. Typically, the CRSS values and hardening parameters (for slip and/or twin systems) are varied in the model calculations by trial and error to best match the model lattice strains vs. applied stress relationship with analogous experimental data. The comparison is made for different *hkl* reflections (i.e., different groups of grains), but only for the two directions of the scattering vector. At the same time, the model and experimental stress-strain dependences are compared. This method was used to study the same extruded magnesium alloy AZ31, initially using the EPSC (elastic-plastic self-consistent) model [60,61] and then the EVPSC (elastic-viscoplastic self-consistent) model [62]. The lattice strains were measured *in situ* in the direction of the applied load and in the transverse direction during tensile and compression tests conducted in the direction of extrusion. As a result, different CRSS values were obtained if different types of models (EPSC or EVPSC [60,61]) or different approximations of the twinning process (in EPSC model in [60–62]) were used. This means that the results of the analysis of experimental data depend on the model applied for interpretation. It should also be emphasised that in calculations, the initial CRSS values and their subsequent changes are usually described by different laws, e.g. the Voce law [38,61,63,67] or the hardening law based on dislocation density [68,69], which can also affect the results obtained by the trial-and-error method.

Another problem with the trial-and-error method is the number of model parameters that are changed in order to fit the calculation results to the experimental data. This is especially important when not all CRSS and hardening parameters are known, and lattice strains are measured in only two directions. Our experience in such data processing has shown that the results obtained are not always unique. Some progress in this aspect was proposed in [17], where a model with a highly reduced number of parameters was used.

The above remarks lead to the conclusion that new methods of CRSS determination directly from diffraction measurements deserve attention. The lack of such methods that would give representative and unambiguous results for a large number of small polycrystalline grains (with sizes of a dozen μm or smaller) is the motivation for this work. The aim of the work is to develop a diffraction method based on the TOF neutron diffraction technique, enabling direct experimental determination (without the use of a model) of unambiguous CRSS values. These experimental CRSSs can be then used in the EPSC model to determine the hardening parameters (by the trial-and-error method). In this way, the ambiguity of the obtained results and their dependence on the theoretical assumptions of the model is avoided or significantly reduced.

The novel methodology for CRSS determination is based on the crystallite group method (CGM), which can be used to determine stresses for groups of grains having a given orientation. Such an experiment seems to be very promising due to the high statistics of the grains for which the measurement is performed. For the first time, CGM was used for the magnesium alloy AZ31 in [18], where the stress evolution was measured for four selected crystal orientations during an *in situ* tensile test. As a result, the CRSS value was determined for the basal system. The purpose of this work is to determine the CRSS value for non-basal slip systems and for the twinning process. The values determined in this way will constitute reliable basic data determined for a representative number of polycrystalline grains. The main advantage of the method developed in this work is that the obtained CRSS values are determined in an unambiguous manner together with the measurement uncertainties.

Another important goal of this work is to determine, for the first time experimentally, stresses in grains with selected orientations in order to explain the huge difference in the mechanical stress-strain diagrams obtained for various load tests performed in different directions on a sample with a strong texture. Moreover, the measured partitioning of the load between grains will allow us to verify the intergranular interactions in the elastic-plastic model, which can differ from those assumed in self-consistent models (Eshelby type of interaction [70]), as shown for example in [71], in the case of pearlitic steel. The verified and modified elastoplastic self-consistent (EPSC) model will be used to predict the evolution of crystallographic texture during plastic deformation, which will be compared with experimental data.

## 2. Experimental methods and material characterization

### 2.1 Material

The studied material was hot-rolled magnesium alloy AZ31, the chemical composition of which is shown in Table 1. The alloy orientation map and crystallographic texture, determined by the EBSD (Electron backscatter diffraction) technique, are presented in Fig. 1; a strong basal texture component is visible. It was found that the average grain size of the alloy is 12.2 µm with a standard deviation of 7.9 µm. An example diffractogram of the initial undeformed material and of the sample loaded with a stress of $\Sigma_{RD} = 247$ MPa in the rolling direction (RD) are shown in Fig. 2.

To measure the stress components using neutron diffraction, three loading experiments were performed; tensile in the rolling direction (later called RDT) [18], compression in the normal direction (NDC), and compression in the rolling direction (RDC).

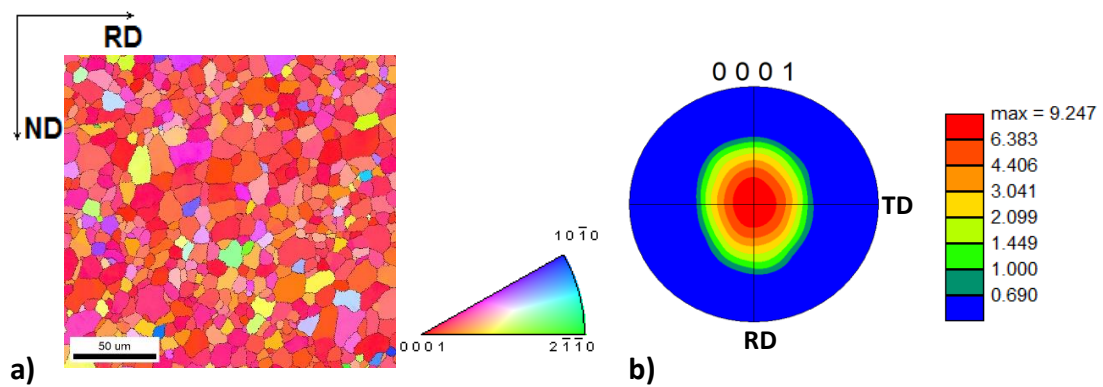

**Fig. 1.** EBSD orientation map (a) and (0001) pole figure for the initial undeformed sample (b).

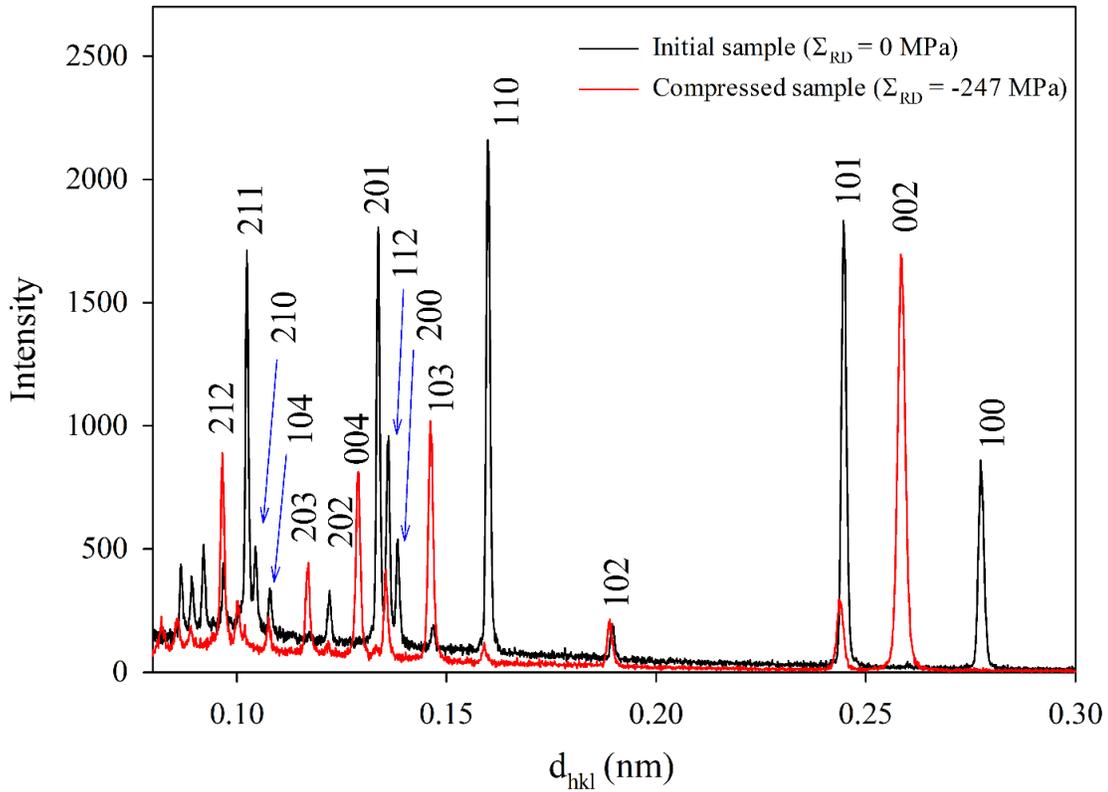

**Fig. 2.** Comparison of the diffractograms for the initial undeformed AZ31 magnesium sample and the sample under compressive load ($\Sigma_{RD} = -247$ MPa) in the rolling direction (RD), measured on the EPSILON - MDS diffractometer (see Fig. 5). The scattering vector was parallel to the RD. A change in texture is visible, especially in the 002 reflection, resulting from the appearance of a tensile twin orientation.

*Ex situ* experiments (before neutron diffraction) were done first, using the same stress rig and the same conditions as during *in situ* neutron measurements, to precisely plan the distribution of measuring steps.

**Table 1.** Chemical composition of magnesium alloy AZ31 according to material specification.

| Element | Al | Zn | Mn | Cu | Mg |
|---|---|---|---|---|---|
| Composition (wt %) | 2.5-3.5 | 0.7-1.3 | 0.2-1.0 | 0.05 | 94.15-96.55 |

### 2.2 Crystallite group method

The experimental methodology used in this work allows the determination of the CRSS values based on the measurement of stresses performed for large grain statistics by neutron diffraction. For this purpose, the crystallite group method (CGM) was used [13,18,57,58] and as shown in the data analysis below, for the first time CRSS values can be determined together with the estimated uncertainties.

The crystallite group method is based on the measurement of the average stress for groups of grains having similar lattice orientations. The measurement of lattice strains is performed in different directions of the scattering vector (along which the lattice strain is determined) using different reflections *hkl*. The idea is to choose such experimental conditions for which the lattice strains are measured for the selected group of grain. Therefore, the scattering vector is perpendicular to the crystallographic plane *(hkl)* corresponding to the lattice orientation of this group, i.e., the measurements are done at the so-called poles $P(\psi,\phi)_{hkl}$ representing normals to the planes *(hkl)*. The orientations of poles $P(\psi,\phi)_{hkl}$ can be presented on a pole figure as shown in Fig. 5 in [18] and Figs. 5 and 7 in the present paper.

The experimental realization of the conditions described above depends on the technique used. In [18], the method of angular dispersion with monochromatic radiation was used, and in this case, the selection of the appropriate pole $P(\psi,\phi)_{hkl}$ is carried out by:

- selection of scattering angle $2\theta$ to satisfy the Bragg relation for a given *hkl* reflection coming from the *(hkl)* plane and corresponding to the chosen pole $P(\psi,\phi)_{hkl}$,
- alignment of the normal to selected *(hkl)* plane along scattering vector by changing sample orientation relative to the laboratory (e.g., the appropriate $\psi$ and $\phi$ angles corresponding to $P(\psi,\phi)_{hkl}$ pole are set using the Eulerian cradle).

For the energy-dispersive diffraction method, e.g., the time-of-flight (TOF) technique used in this work, experimental data for multiple *hkl* reflections are available simultaneously for a given orientation of the scattering vector and constant scattering angle $2\theta$ (due to a continuous spectrum of wavelength). However, in the used equipment it is not possible to change the orientation of the sample with respect to scattering vector. Therefore, the selection of an appropriate pole $P(\psi,\phi)_{hkl}$ must be carried out through:

- selection of the reflection *hkl* from the available diffraction pattern which corresponds to the given *(hkl)* plane for the chosen pole $P(\psi,\phi)_{hkl}$ (all available reflections are measured with the same constant orientation of the scattering vector characterised by given $\psi$ and $\phi$ angles),
- if more detectors are available the selection of appropriate reflection *hkl* is repeated for each detector (i.e., for different orientations of the scattering vector).

Therefore, the difference between the two techniques consists in the selection of *hkl* reflections that satisfy the Bragg condition for the selected *(hkl)* plane. It should also be emphasised that the angular dispersion method (with the Eulerian cradle) allows for a strict selection of the set of $P(\psi,\phi)_{hkl}$ poles corresponding to the orientation of the lattice; however, the lattice strains corresponding to the given lattice orientations are measured consecutively and not at the same time. This leads to long measurements during which some (usually small) relaxation of the applied macrostress may occur during plastic deformation. On the contrary, the dispersion energy method used in this work allowed the measurement of lattice strains for many poles $P(\psi,\phi)_{hkl}$ simultaneously thanks to the use of a continuous spectrum of wavelength and nine detectors with different orientations relative to the incident beam (scattering vector $2\theta = 90°$ for all detectors, details are given below). In this way, nine scattering vector orientations are available for a sample fixed in a tensile/compression rig, and each detector simultaneously measures the diffraction pattern with different *hkl* reflections. Then, each diffractogram is analysed to find reflections from planes *(hkl)* corresponding to the given lattice orientations. However, it should be emphasised that the directions of the scattering vectors are usually not strictly parallel to the normals to the selected planes *(hkl)*. When selecting the poles, the criterion was adopted that the difference between the normal to the plane *(hkl)* and the scattering vector should not exceed 5º.

### 2.3 Experiments

To determine the stresses for given groups of grains with selected orientations (Figs. 3 and 4), *in situ* neutron diffraction measurements of lattice strains were performed for samples subjected to various uniaxial loads. The first experiment (RDT) was carried out on a TKSN 400 (HK9) diffractometer at the Nuclear Physics Institute in Řež (Czech Republic) [72]. Due to the availability of a monochromatic beam with a wavelength of λ=1.1580 Å, the measurements were made using the angle dispersive method. To perform lattice strains measurements for the selected orientations

A, B, C and D (see Fig. 3 and Fig. 4), the tensile rig was mounted in the Eulerian cradle (Fig. 5a). The sample was a 2 cm long bar with a square cross section of 4 mm per side. The experimental setup used at the TKSN 400 diffractometer is shown in Fig. 5, where the orientation of the scattering vector $\overrightarrow{\Delta k}$ relative to the $x_i$ coordinate system (in which the stress tensors $\sigma_{ij}$ are defined) is determined through the angles $\psi$ and $\varphi$ (being changed by Eulerian cradle rotations). The coordinates of this orientation in the pole figure are shown in Fig. 5b. The results of these measurements have already been published [18] and reprocessed data are used in this work to complete the analysis of new RDC and NDC experiments.

It should be emphasised that in all experiments presented in this paper (RDT, RDC and NDC), the $x_3$ axis corresponding to the central point of the pole figure was chosen as parallel to the direction of the applied load, which simplifies both the analysis and the presentation of the results. The orientations of the sample directions RD, TD and ND (rolling direction, transverse direction and normal direction) in relation to the $x_i$ axes are given in Table 2. The coordinates of poles for different lattice plane normals corresponding to the orientations A, B, C and D selected in the RDT experiment for lattice strains measurements are presented in Table A1.1 (see Appendix 1).

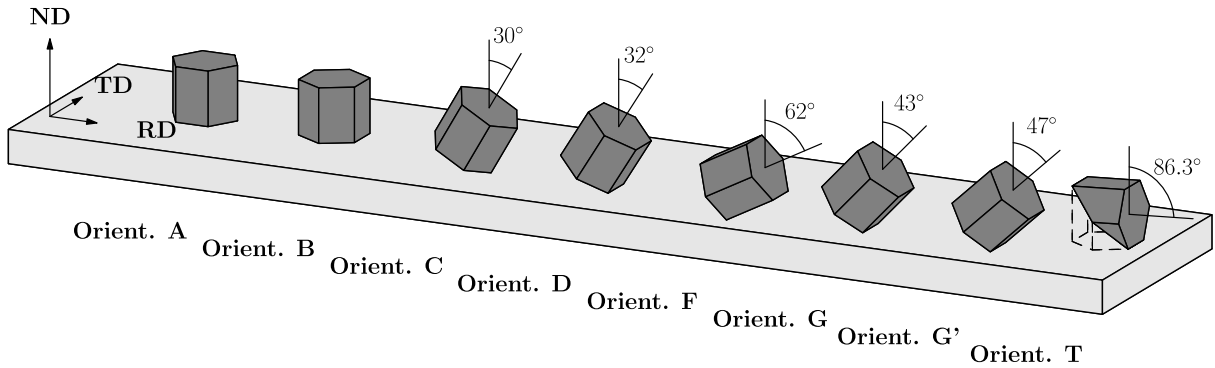

**Fig. 3** Visualisation of orientations examined during all experiments.

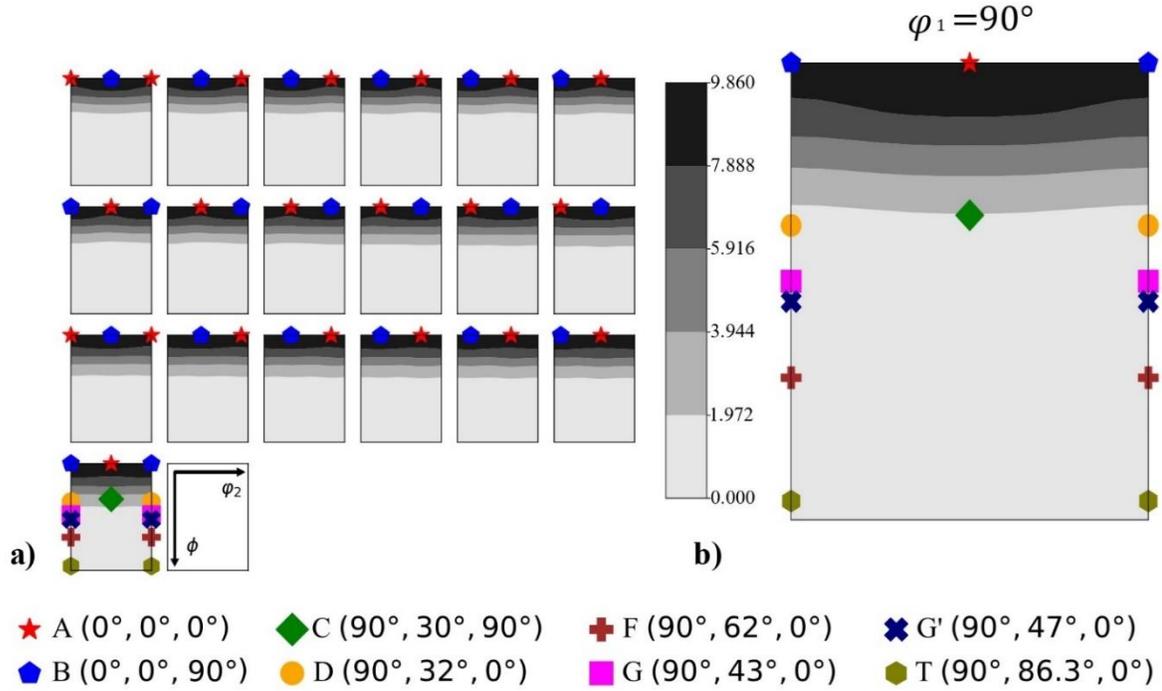

★ A (0°, 0°, 0°)    ◆ C (90°, 30°, 90°)    ✚ F (90°, 62°, 0°)    ✖ G' (90°, 47°, 0°)
● B (0°, 0°, 90°)    ● D (90°, 32°, 0°)    ■ G (90°, 43°, 0°)    ● T (90°, 86.3°, 0°)

**Fig. 4** Experimental ODF obtained from EBSD measurements for the initial undeformed sample. The crystal orientations studied in this work are marked: (a) on full ODF graph, (b) on slice for $\varphi_1=90°$.

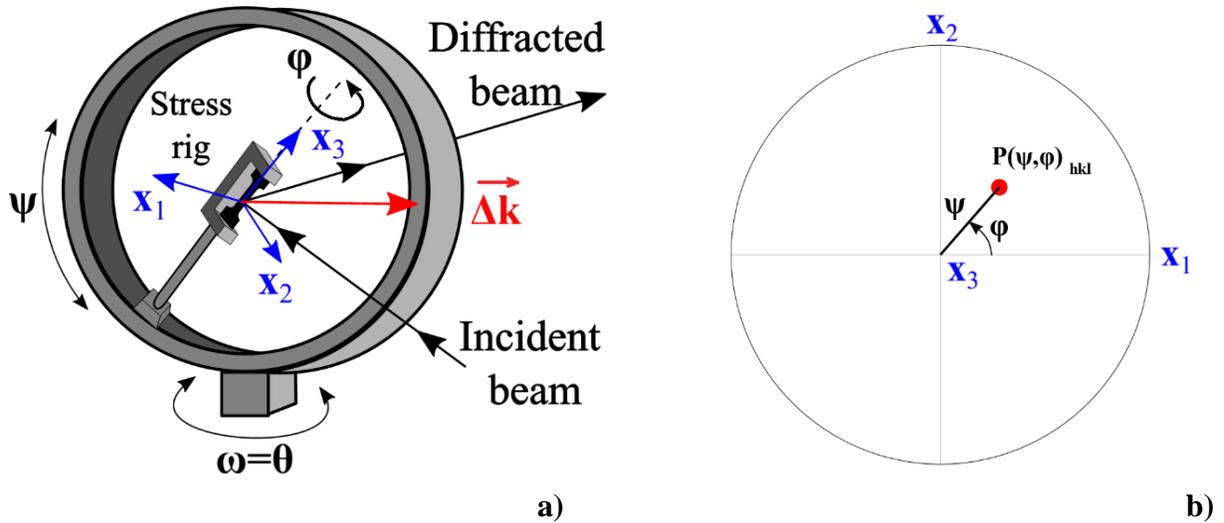

**Fig. 5.** (a) Scheme of Eulerian cradle used on TKSN 400 (HK9) diffractometer to change the orientation of the sample with respect to the scattering vector $\vec{\Delta k}$ by changing angles $\psi$ and $\varphi$, while reflection $hkl$ is chosen by setting scattering angle $2\theta$. (b) Selected pole $P(\psi,\phi)_{hkl}$ in pole figure for which the lattice strain is measured in the crystallite group method.

**Table 2.** Orientations of $x_i$ axes with respect to sample directions for non-rotated sample ($\alpha = 0°$).

| Experiment | RDT and RDC | NDC |
|---|---|---|
| Orientations of $x_i$ axes | $x_1\|\|$TD, $x_2\|\|$ND, $x_3\|\|$RD | $x_1\|\|$TD, $x_2\|\|$RD, $x_3\|\|$ND |

The next two experiments (NDC and RDC) were performed on the EPSILON-MDS (Multi Detector System) diffractometer at the Frank Laboratory for Neutron Physics (FLNP) in the Joint Institute for Nuclear Research (JINR) in Dubna (Russia) [73]. The time diffractometer is placed at the IBR-2 pulse reactor in the JINR and uses the time-of-flight (TOF) method. Samples measured during *in situ* experiments had a cylindrical shape and their initial dimensions were 24.04 mm length and 13.9 mm diameter in the NDC experiment and 27.4 mm length and 13.9 mm diameter in the RDC experiment. In this case, the energy dispersive method was used, and the lattice strains were measured in the directions defined by nine detectors having given orientations with respect to the sample mounted in the compression rig. The measurement geometry of the detector groups of this instrument is presented in Fig. 6. In both experiments (NDC and RDC), the same orientation of the coordinate system $x_i$ with respect to instrument was used for stress tensor ($\sigma_{ij}$) definition, while the orientations of the RD, ND, and ND sample directions in relation to $x_i$ coordinates were different for NDC and RDC measurements (see Table 2).

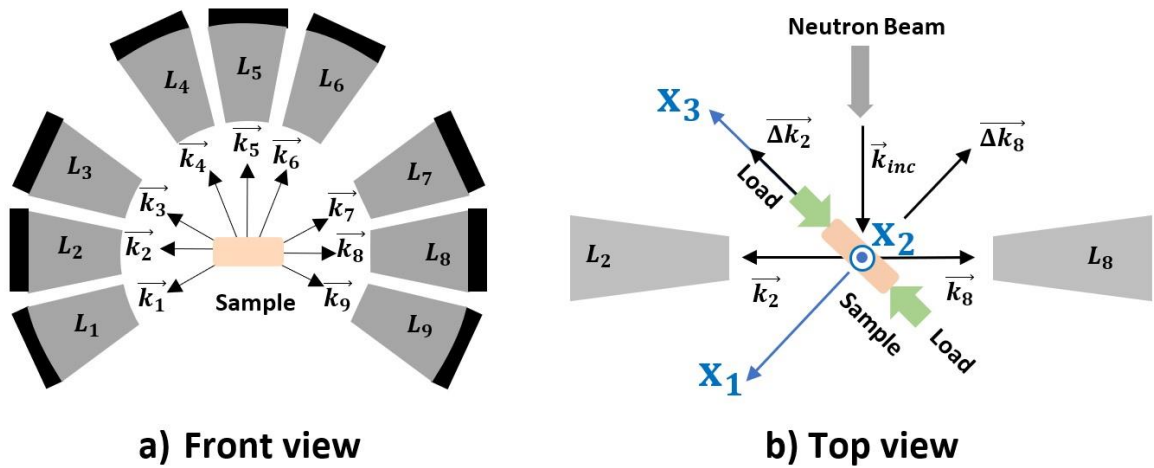

a) Front view    b) Top view

**Fig. 6.** Schematic diagram of detector banks at the EPSILON-MSD diffractometer showing in figure (a) the directions of the diffracted beams denoted by wave vectors $\vec{k}_i$ (for each vector the incident beam is perpendicular to $\vec{k}_i$, i.e., $2\theta=90°$). Two example scattering vectors $\overrightarrow{\Delta k_2}$ and $\overrightarrow{\Delta k_8}$, shown in figure (b), are placed on the plane containing the incident wave vector $\vec{k}_{inc}$ and diffracted wave vectors $\vec{k}_2$ and $\vec{k}_8$, respectively. The system $x_i$ is defined by the unit vectors $\hat{x}_i$ given by the relations: $\hat{x}_3 = \overrightarrow{\Delta k_2}/|\overrightarrow{\Delta k_2}|$, $\hat{x}_1 = -\overrightarrow{\Delta k_8}/|\overrightarrow{\Delta k_8}|$ and $\hat{x}_2 = \hat{x}_3 \times \hat{x}_1$.

The compression rig allowed the cylindrical sample to be rotated by any angle $\alpha$ about its axis (parallel to $x_3$), which increased the number of measured orientations. One turn of $\alpha = 90°$ (which brings the ND direction to the $x_1$ axis) was done during the RDC experiment. In addition, the combination of detector geometry and sample symmetry enabled measurements for multiple poles, corresponding to many orientations used in further analysis. In the NDC experiment, the orientations measured were A, B, D, F and G, while in the RDC experiment these were A, B, D, F and G'. These orientations are marked on the orientation distribution function (ODF, see Fig. 4) and their visualisation is showed in Fig. 3. The coordinates of poles $P(\psi, \phi)_{hkl}$ for different lattice plane normals used in the measurements of lattice strains in these experiments are listed in Table A1.1 (see Appendix 1). Examples of orientations used in the NDC and RDC experiments are also presented by pole figures in Fig. 7, where the positions of scattering vectors corresponding to the detectors and their symmetrical counterparts are marked, considering the symmetries resulting from the crystallographic texture of the sample (the load is in the centre of each pole figure). It should be emphasised that the positions of the scattering vector (or equivalent) do not exactly coincide with the poles, therefore only the closest position (discrepancy less than 5°) was selected for the analysis.

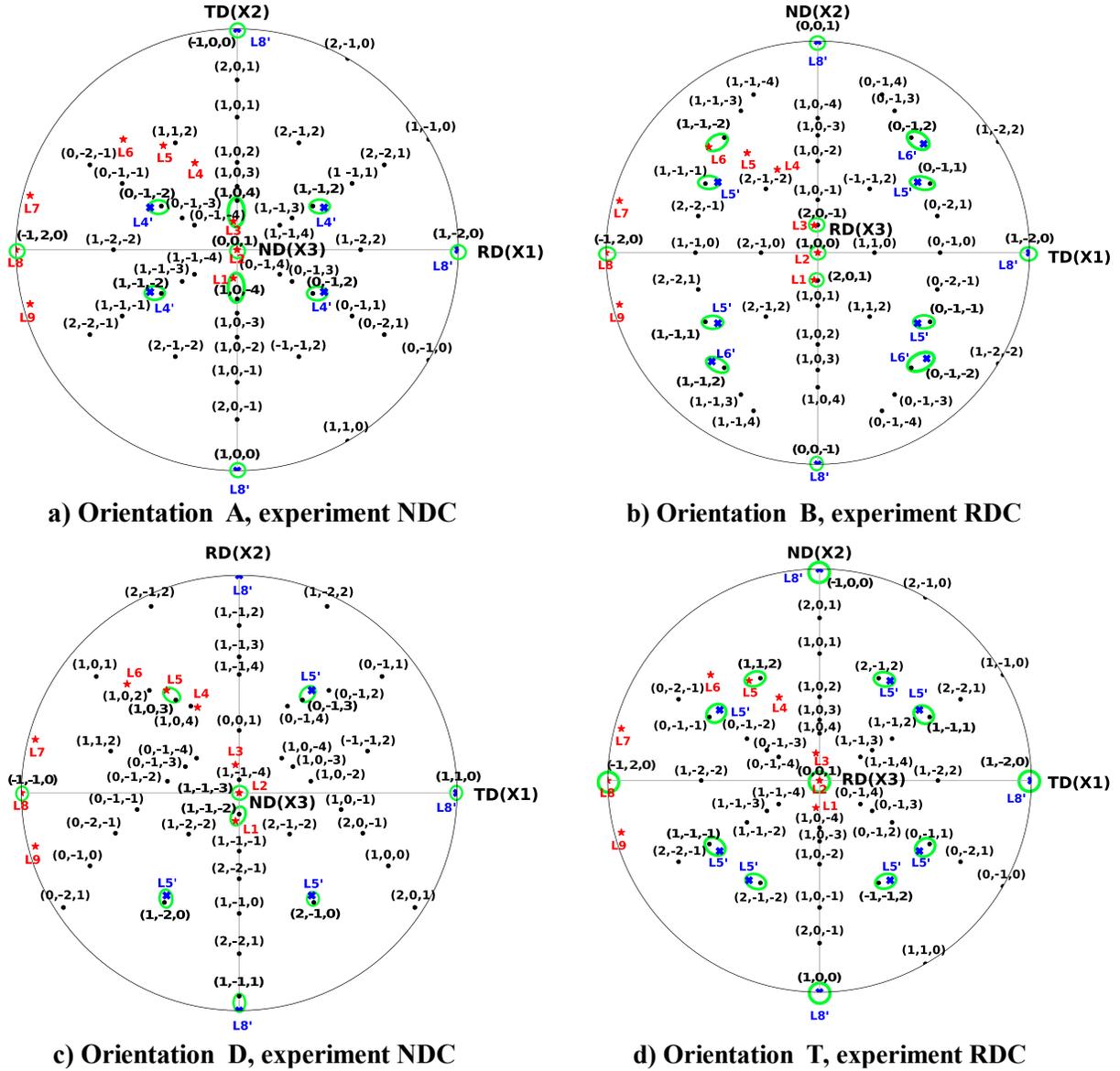

**Fig. 7.** Poles $P(\psi,\phi)_{hkl}$ used to determine the stresses for different grain orientations and experiments are marked with black dots. The orientations of the scattering vectors corresponding to the nine detectors of the EPSILON-MDS diffractometer are indicated by the red stars L1-L9, while the blue stars L1'-L9' are used for the equivalent position determined by texture symmetry and sample rotation during the experiment. The poles assigned to the positions of the scattering vector or equivalent are in the green circles.

During each experiment, the force was applied until the specified stress value was obtained, after which the rig was stopped to maintain a constant sample strain. When entering the range of plastic deformations, after each load increase, the macroscopic stresses determined from the force sensor decreased until stabilization; therefore, diffraction measurements were postponed by 30 minutes.

Macroscopic stress-strain curves obtained for each of the experiments determined at constant macrostrain as the average of the values measured at the beginning and end of the diffraction measurement are presented in Fig. 8. These averages directly correspond to the lattice strains determined *in situ* in diffraction measurement. It is worth noting that there is a large difference in the mechanical behaviour of the sample depending on the direction and sign of the applied load. During the RDC experiment, a plateau occurred during which the twins were created.

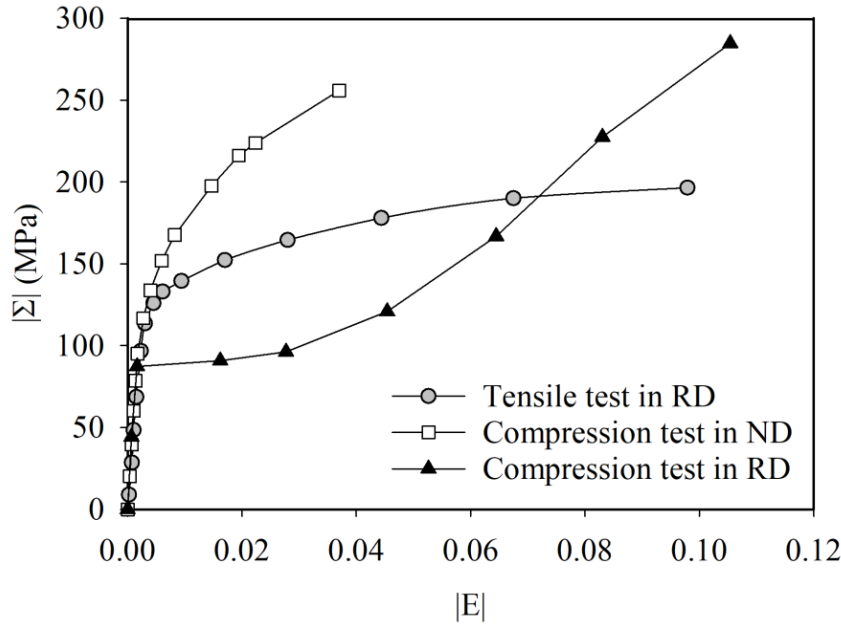

**Fig. 8.** Results of three different tests conducted for the tested alloy AZ31. Each value of the macrostress applied - force divided by the cross section of the sample.

## 3. Direct determination of grain stresses and CRSS values

### 3.1 Determination of stresses for a given orientation

The important goal of this work is to determine experimentally the stresses for selected groups of crystallites, both those which are very numerous due to crystallographic texture, but also those which are very few and do not play a significant role during the deformation process. The stress evolution for a given grain orientation was determined using the crystallite group method [13,18,57,58], i.e. from the set of lattice strains measured during a tensile/compression test for appropriate *hkl* reflections and orientations of the scattering vector, corresponding to the poles assigned to the tested grains – see Table A1.1 (see Appendix 1). It should be emphasised that for a given reflection *hkl*, the lattice deformation is measured in the direction of the scattering vector, which is normal to the crystallographic planes *(hkl)*. The orientation of the scattering vector is determined by the angles $\psi$ and $\varphi$. The stresses $\sigma_{ij}^{CR}$ for selected groups of crystallites were determined from experimental lattice strains, using the least square method [74], based on the relation:

$$<\varepsilon(\varphi,\psi)>_{hkl} = \frac{<d>_{hkl}-<d>^0_{hkl}}{<d>^0_{hkl}} = F_{ij}^R(hkl,\varphi,\psi)\,\sigma_{ij}^{CR} \qquad (1)$$

Where the $\sigma_{ij}^{CR}$ tensor is defined always with respect to the $x_i$ system of coordinates (see Figs. 5 and 6), and $<d>_{hkl}$ and $<d>^0_{hkl}$ are the interplanar spacings measured in the direction of the scattering vector (given by angles $\psi$ and $\varphi$) using *hkl* reflection for the sample under applied load and for the initial non-loaded sample, respectively. $F_{ij}^R(hkl,\varphi,\psi)$ are the so-called X-ray stress factors or diffraction elastic constants calculated by the Reuss approach [13,43,75,76], which relate the lattice strains measured for a selected group of crystallites with the stress tensor $\sigma_{ij}^{CR}$ for this group (assuming the same $\sigma_{ij}^{CR}$ for all grains in the group).

It should be emphasised that lattice strains for twins cannot be determined directly from the measurements due to the lack of twin orientations in the initial non-deformed sample. Therefore, the values of $<d>^0_{hkl}$ (needed in Eq. 1) for twin orientations were recalculated from an interpolation method that used the interplanar spacing available in the initial sample and measured with different detectors. In the calculations presented in Appendix 2a, the standard uncertainties of the values $<d>^0_{hkl}$ determined in this way were estimated. This approach was previously used in [61].

The uncertainty analysis of the stresses determined in this work is described in detail in [15] and Appendix 2b, and is based on the least squares method given in [74].

The $F_{ij}^R(hkl,\varphi,\psi)$ factors were calculated from the ODF presented in Fig. 4 and the single crystal elastic constants: $C_{11} = 59.3$ GPa, $C_{33} = 61.5$ GPa, $C_{44} = 16.4$ GPa, $C_{12} = 25.7$ GPa, $C_{13} = 21.4$ GPa [77], using the Reuss model [13,43], considered as the proper approach for the groups of studied crystallites [76]. In this model, the same stress is assumed for grains belonging to a given group. It should also be emphasized that the choice of coefficients $F_{ij}^R(hkl,\varphi,\psi)$ or single crystal elastic constants (as in [13,18,57,58]) in the data analysis is irrelevant due to the almost isotropic properties of Mg crystals [78], leading to negligible differences obtained for both choices.

The first analysed experiment was the tensile test performed in the rolling direction (RD). In Fig. 9, the grain stresses for orientations A, B, C and D versus the macroscopic stress corresponding to the applied load are shown. The sets of strains used for stress determination are shown in [18] and in Table A1.1 (Appendix 1). Because of the sample and texture symmetry, only the non-zero principal stresses $\sigma_{ii}^A$ and $\sigma_{ii}^B$ were assumed for the A and B orientations. In the case of the orientation C, the non-zero $\sigma_{23}^C$ component is also shown (defined with respect to coordinate system $x_i$). This shear stress can be different from zero because the crystallite axis <c> is tilted from the $x_2$ axis (normal to the surface of the sample, i.e., ND) toward the direction $x_3$ (RD) along which the tensile force was applied (see Fig. 3 and Table 2). Therefore, these axes of the stress tensor defined with respect to the sample system may not coincide with the principal stress axes. Finally, in the case of D orientation only the $\sigma_{33}^D$ component was determined because the strains were measured only in two directions. Additionally, the results of the EPSC model are shown in Fig. 9 (they are discussed in Section 4.2).

Analysing the experimentally determined components of stress tensor, it can be noticed that in the beginning of compression test ($\Sigma_{RD}$ smaller than about 70-90 MPa) the grain stresses are practically equal to the applied load, i.e. $\sigma_{33}^{A,B,C,D} = \Sigma_{RD}$ and other stress components are equal to zero (see Fig. 9). This means that, due to the low elastic anisotropy of crystallites, the stresses at all grains having different lattice orientations are almost equal. However, when plastic deformation begins ($\Sigma_{RD}$ between 70-90 MPa), the partitioning of stresses between grains starts. A higher tensile stress in the loading direction is observed at grains having orientation A and B (with the <c> axis parallel to the ND, i.e., $\sigma_{33}^{A,B} > \Sigma_{RD}$). Simultaneously, the value of $\sigma_{11}^{A,B}$ remains almost equal to zero, while a small compressive stress, $\sigma_{22}^{A,B} < 0$, is generated. The deviation of the grain stress components from the macroscopic values increases with increasing applied load. The opposite behaviour is demonstrated by the "tilted grains" with orientations C and D, for which the tensile stress in the direction of the load decreases comparing to the macroscopic value ($\sigma_{33}^{C,D} < \Sigma_{RD}$). This means that plastic deformation occurred for the grains D and C, causing transfer of a part of the load to other grains that remain elastic. Representatives of the still elastically deformed grains are those with the A and B orientations (for such grains, the stress $\sigma_{33}^{A,B}$ is greater than the macroscopic stress $\Sigma_{RD}$). The interaction between grains in the direction perpendicular to the load is more complex than in the loading direction and it could be explained for example by the EPSC model.

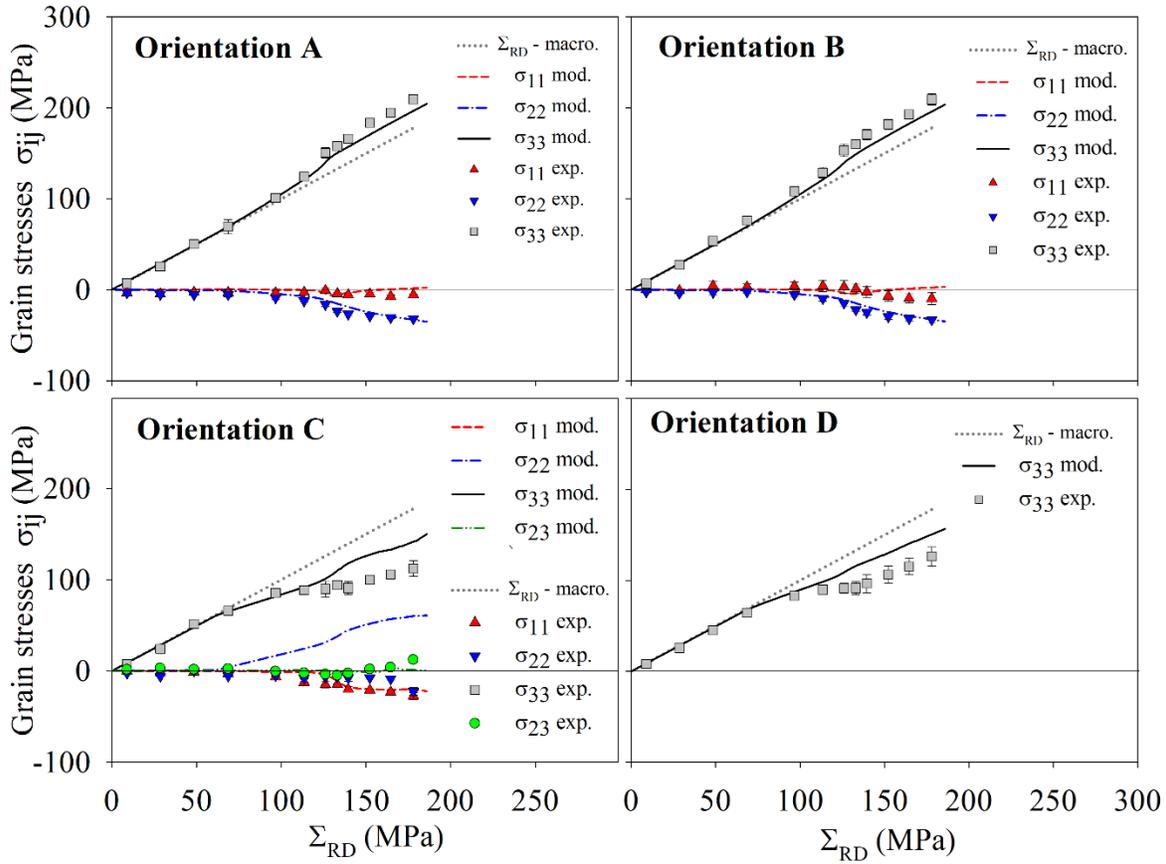

**Fig. 9.** Evolution of grain stresses for the orientations A, B, C and D versus macrostress $\Sigma_{RD}$ during the tensile test performed along the RD compared with the model using threshold approximation (see Section 4). The evolution of macroscopic stress $\Sigma_{RD}$ is drawn with a dotted line. The error bars correspond to standard uncertainties calculated according to Appendix 2b.

Subsequently, it can be noticed that for $\Sigma_{RD} > 125$ MPa the trend of the plots obtained for the orientations A and B again changes, i.e., the deviation of the grain stresses from the macroscopic values stabilizes and does not increase anymore. This can be interpreted as the beginning of plastic deformation for the grains A and B. At the same time, a change in the behaviour of stresses also takes place for the grains C and D, i.e., greater stress is transferred to these grains in response to the plastic deformation of the grains with orientations such as A and B. It should be emphasised that the $\Sigma_{RD}$ thresholds, where the trends of plots change, cannot be directly used in Schmid's law to determine RSS values for activated slip systems because of the occurrence of intergranular stresses. Therefore, in this work the calculation of the RSS is based on the stress tensor components determined for selected orientations using neutron diffraction.

In the second and the third experiments, the sample was compressed in the ND and RD directions, respectively. The set of reflections used for lattice strain measurements is shown in Table A1.1 (see Appendix 1) and it was assumed that due to the sample and crystal symmetry, only the non-zero principal stresses $\sigma_{ii}^A$ were determined for the orientation A, while in the case of orientations D, F, G and G' the non-zero $\sigma_{23}^{D,F,G,G'}$ component was additionally determined. Similarly, as in the previous experiment, this shear stress can be different from zero because the crystallite axis <c> is tilted from the $x_3$ (ND) sample axis (along which the compressive force was applied) towards the direction $x_2$ (RD). The results of stress component evolution during the NDC experiment are presented in Fig. 10 for orientations A, D, F and G (the result for orientation B was almost identical with the one for orientation A, due to texture and crystal symmetry). Similarly as in Fig. 9, the results of the EPSC model are shown but are analysed in Section 4.2.

Analogically as in the previous experiment, it can be noticed that at the beginning of the test (for $|\Sigma_{ND}|$ smaller than about 70-90 MPa) the grain stresses are practically equal to the applied load, i.e. $\sigma_{33}^{A,B} = \Sigma_{ND}$, and the other stress components are equal to zero (cf. Fig. 10). This confirms that the crystallites exhibit low crystal anisotropy leading to the same value of stress for all grains having different lattice orientations. However, when plastic deformation starts (at about $|\Sigma_{ND}| \approx$ 70 – 90 MPa), greater compressive stresses in the loading direction are observed at grains having orientation A and B (i.e., $|\sigma_{33}^{A,B}| > |\Sigma_{ND}|$), and simultaneously non-zero tensile stress tensor components $\sigma_{11}^{A,B} \approx \sigma_{22}^{A,B} > 0$ are generated at the transverse directions. The deviation of the grain stress components from the macroscopic values increases with increasing load. The opposite behaviour is demonstrated by the "tilted grains" with orientations D, F and G, for which the compressive stress in the direction of the load decreases comparing to the macroscopic value ($|\sigma_{33}^{D,F,G}| < |\Sigma_{ND}|$). Additionally, for these grains the non-zero stresses are generated in the directions perpendicular to the load, but their behaviour is complex. Similarly, as in the previous experiment, transfer of the stress from plastically deformed grains (e.g., D, F and G orientations) to elastically deformed ones (e.g., orientations A and B) occurred. Subsequently, it can be noticed that for $|\Sigma_{ND}| > 180$ MPa the trend of the plots obtained for the orientations A and B changes again, and the deviation of the grain stresses from the macroscopic values does not increase. This means the beginning of plastic deformation for the grains A and B. The changes in the behaviours of stresses at grains D, F and G are not significant for $|\Sigma_{ND}| > 180$ MPa.

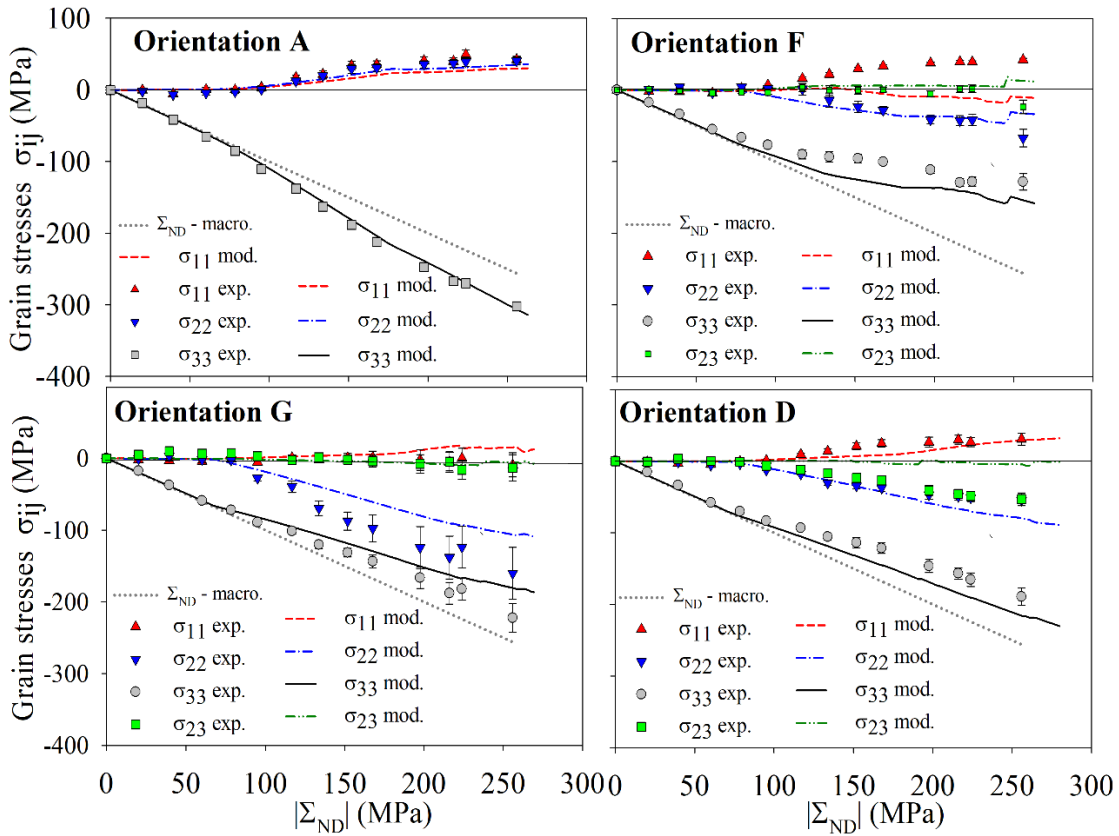

**Fig. 10.** Evolution of the grain stresses for orientations A, D, F and G versus macrostress $|\Sigma_{ND}|$ during the compression test performed along the ND compared with a model using threshold approximation (see Section 4). The evolution of macroscopic stress $\Sigma_{ND}$ is drawn with a dotted line. The error bars correspond to standard uncertainties calculated according to Appendix 2b.

In the third experiment, a compressive load in the direction of the RD was applied, and in this case the twinning process took place at about $|\Sigma_{RD}| \approx 90$ MPa for a significant number of grain orientations. Before the twinning process, insignificant deviations of the stresses from the macroscopic stress were observed for the studied orientations A, B and G' (cf. Fig.11).

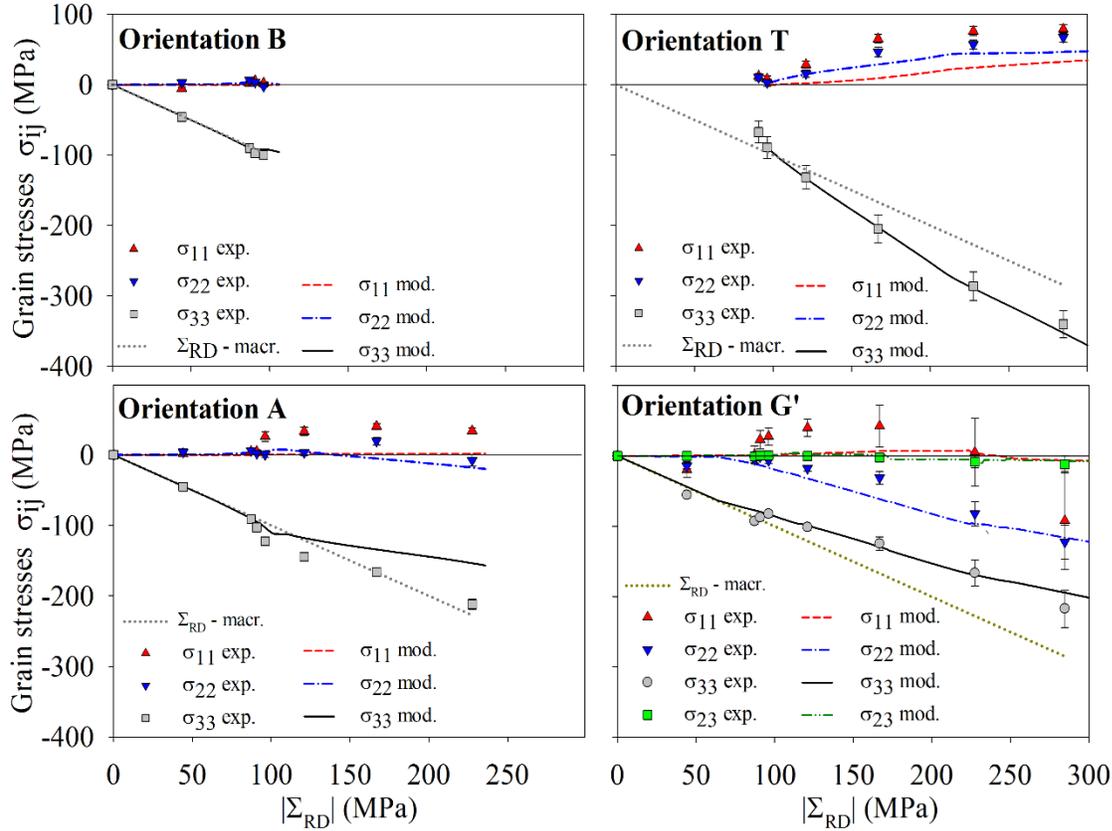

**Fig. 11.** Evolution of the grain stresses for orientations A, B, T and G' versus macrostress $|\Sigma_{RD}|$ during the compression test performed along the RD compared with a model using threshold approximation. The grains having B orientation are transformed to twins (T-orientation) at approximately $|\Sigma_{RD}|$=90 MPa. The evolution of macroscopic stress $\Sigma_{RD}$ is drawn with a dotted line. Very similar results were obtained for the continuous approximation (for details concerning the model see Section 4). The error bars correspond to standard uncertainties calculated according to Appendix 2b.

During twinning occurring at approximately constant load, the twins (orientation T) are formed mostly from crystallites having B orientation, which completely disappeared. The new twins, when they are born, exhibit compressive stress $\sigma_{33}^{T}$, for which the absolute value $|\sigma_{33}^{T}|$ is slightly smaller than the absolute value of macrostress $|\Sigma_{RD}|$, but the value $|\sigma_{33}^{T}|$ rises along with increasing of the load and quickly exceeds the value $|\Sigma_{RD}|$. Subsequently, the deviation of the $|\sigma_{33}^{T}|$ from $|\Sigma_{RD}|$ value increases until the $|\Sigma_{RD}|$ reaches about 220 MPa, where the deviation stabilizes. The transverse stresses $\sigma_{11}^T$ and $\sigma_{22}^T$ are tensile; they start from zero value and progressively increase until $|\Sigma_{RD}| \approx 220$ MPa. This behaviour is characteristic for the elastically deformed grains in the plastically deformed material, i.e., elastic grains accumulate higher stresses as compared to macrostress. The A-oriented grains also transform into twins; this transition is slower and some of them persist longer compared to the B-oriented grains. The value $|\sigma_{33}^A|$ for the A grains is higher than $|\Sigma_{RD}|$ at the beginning, but with increasing load, the $|\sigma_{33}^A|$ approaches the macroscopic value $|\Sigma_{RD}|$, while the stresses in the direction perpendicular to the load are small and tensile or they are close to zero. The behaviour of the G'- oriented grains is similar to that observed in the NDC experiment, i.e., the absolute value in the load direction $|\sigma_{33}^{G'}|$ rises more slowly than the macroscopic value $|\Sigma_{RD}|$ when plastic deformation begins.

### 3.2 Determination of CRSS for slip and twinning systems

Having determined the evolution of the $\sigma_{ij}^{CR}$ stresses for grains having different orientations during three different modes of deformation, the resolved shear stresses (RSS) for different slip and twin systems can be calculated. Then by analysing the behaviour of chosen RSS for specific grain orientations during three different tests, the CRSS values for these systems can be found. For this purpose, the following rules to find out the active slip or twin system were applied:

- a potentially active system is one for which the RSS takes the maximum value among all symmetrically equivalent slip systems – this system is activated when its RSS reaches the CRSS value (Schmid criterion),
- when the system is activated, the RSS on this system starts to grow more slowly or does not change significantly – the RSS value for the active system is considered as the CRSS value,
- if the evolution of the RSS on more than one system indicates their activation at the same sample load, another experimental test should be analysed to select the system actually activated,
- the slip systems for which the values of the RSS calculated from measured tensor $\sigma_{ij}^{CR}$ are very low, i.e. less than about 10 MPa (or lower than zero in the case of twin system), are treated as inactive – this concerns systems for which the Schmid factor calculated for the uniaxial applied stress equals zero.

Using the above rules, the analysis of the measurement results was carried out considering the potential possibility of activation of the slip and twin systems during the three conducted experiments when plastic deformation begins. The behaviours of the transverse stresses and shear stress $\sigma_{23}^{G'}$ (which is expected, similarly as in the RDT experiment) for this grain orientation are caused by interaction between grains; therefore, their evolution should be explained by the model (cf. Section 4.2).

- In the case of the tensile test performed in the RD, the basal slip system cannot be activated in the grains having A and B orientations because the load applied to the sample is parallel to the basal plane, leading to zero value of the Schmid factor for macrostress $\Sigma_{RD}$. The RSS on the tensile twin system is negative, therefore twinning cannot occur. The other slip systems can be active for such grain orientations. On the other hand, the basal system and all other systems can be activated for C orientation.

- During the compression test in the ND, the only potentially active system for the orientations A and B are the pyramidal <c+a> systems for which the Schmid factor is not equal to zero. The RSS for twinning is negative, therefore twinning cannot occur. The other systems are not activated due to the zero value of the Schmid factor when uniaxial $\Sigma_{ND}$ stress is perpendicular to the slip plane (as for the basal system), parallel to the slip plane (as for the prismatic system) or perpendicular to slip direction (as for basal, prismatic and pyramidal <a> systems). For the D, F and G orientations, all slip systems including the basal system can be active.

- In the case of the compression test in the RD, the RSS on tensile twin system is positive and significant for the orientations A and B. This gives a chance to find out the value of the CRSS for the twinning process. The RSS evolution for G' orientations are also considered and, in this case, all slip systems can be active. Moreover, for the tensile twins (T-orientation), the Schmid factor calculated for the $\Sigma_{RD}$ stress is close to zero for the basal, prismatic, and pyramidal <a> systems, while for the tensile twin system the RSS is negative. Therefore, in grains having T orientation, only the pyramidal <c+a> systems can be activated.

Considering the above remarks, initially the evolution of the maximum RSS (among all symmetrically equivalent systems) for the basal system in the tilted orientations were analysed. The tilted orientations are those for which the <c> axis is not perpendicular to one of the directions RD, TD or ND and the Schmid factor is not zero for the basal system. A common feature of such orientations is that the basal system can be activated, while for the A, B and T orientations it is inactive due to the zero value of the Schmid factor. It was found that after the onset of plasticity, the grain stress for the tilted orientations is always lower than the macroscopic value, i.e., these grains are softer than e.g. A, B and T orientations (cf. Figs. 9 - 11). The conclusion is that the CRSS and work hardening are the lowest for the basal system compared to other systems. The evolution of the maximum RSS for basal system was shown in Fig. 12 for the tilted orientations D and F during the compression test in the ND. It is clearly visible that after the linear increase in the maximum RSS during elastic deformation, the value of the RSS stabilizes and does not change for a certain range of deformation (or even slightly decreases). The CRSS is equal to the RSS value at the point where the trend of the plot has radically changed, and it is determined as the average of the points closest to the trend change. The uncertainty of the CRSS is estimated based on measurement uncertainties of these points. In the case of the compression test performed along the ND, the mean CRSS value for the base system is 28 MPa (calculated for the D and F orientations) and this value was used in further analysis (cf. Table 3). In the other two experiments, the CRSS values are more difficult to determine.

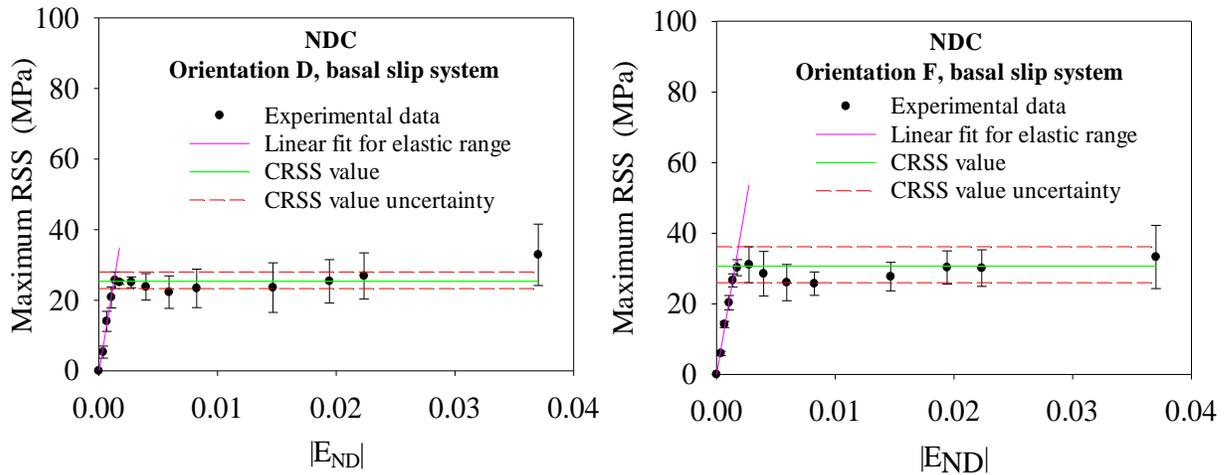

**Fig. 12.** Evolution of the RSS for the basal system (activated in grains having orientations D and F) versus sample strain $|E_{ND}|$ during the compression test performed along the ND. The error bars correspond to standard uncertainties of the RSS (cf. Appendix 2b). Determined CRSS values and their uncertainties are marked with horizontal lines.

Additionally, the values of CRSS for pyramidal systems <c+a> can be determined from the compression test in the ND. In this case, the evolution of maximum RSS on these systems in the grains with orientations A and B were considered. For such orientations, only pyramidal systems <c+a> exhibit non-zero Schmid factors. As shown in Fig. 13, a linear dependence of the maximum RSS versus $|\Sigma_{ND}|$ during elastic deformation of the sample occurred. Subsequently, the RSS increases faster due to the load transfer to those grains from the grains with tilted orientations in which the basal system is activated. Eventually, the upward trend in RSS value suddenly slows down and stabilizes, which means activation of the <c+a> pyramidal systems (cf. Fig. 13). Therefore, the intersection of the regression lines fitted below and above the trend change can be identified as the point at which the RSS value equals the CRSS value. Additionally, the uncertainty of the CRSS value obtained in this way can be estimated by considering the maximum difference between the intersections of the extreme lines corresponding to the uncertainty range for both fitted lines. The values of the CRSS with their uncertainties for the first and the second order pyramidal <c+a> systems activated in grains A and B during the NDC experiment are given in Table 3. It is worth noting that we cannot distinguish which of the two systems (first or second order) is activated, it is also possible that they are activated together.

Similar analysis of the pyramidal <c+a> systems activation can be performed considering the twin grains (T-oriented) during the RDC test. Except for intergranular stresses, the stress state of the T-oriented grains in the RDC test is similar to the A and B-oriented grains in the NDC test, i.e., the load is parallel to the <c> axis. Based on the analysis of the maximum RSS values on the <c+a> systems (both first and second order), it was found that the CRSS values are slightly higher for the T-oriented grains compared to the A and B-oriented grains (cf. Fig. 13e, 13f and Fig. 13a-d), i.e., twins show a higher CRSS value than the grains in the initial (undeformed) sample (cf. Table 3). It should be emphasised that due to the small number of experimental points and their significant uncertainty, the uncertainty of the CRSS value for <c+a> slip systems cannot be exactly estimated for T-oriented twins. Therefore, the uncertainty measure was defined as the RSS interval determined by two experimental points closest to the point where the curve changes its trend (see Fig. 13e, 13f, and Table 3).

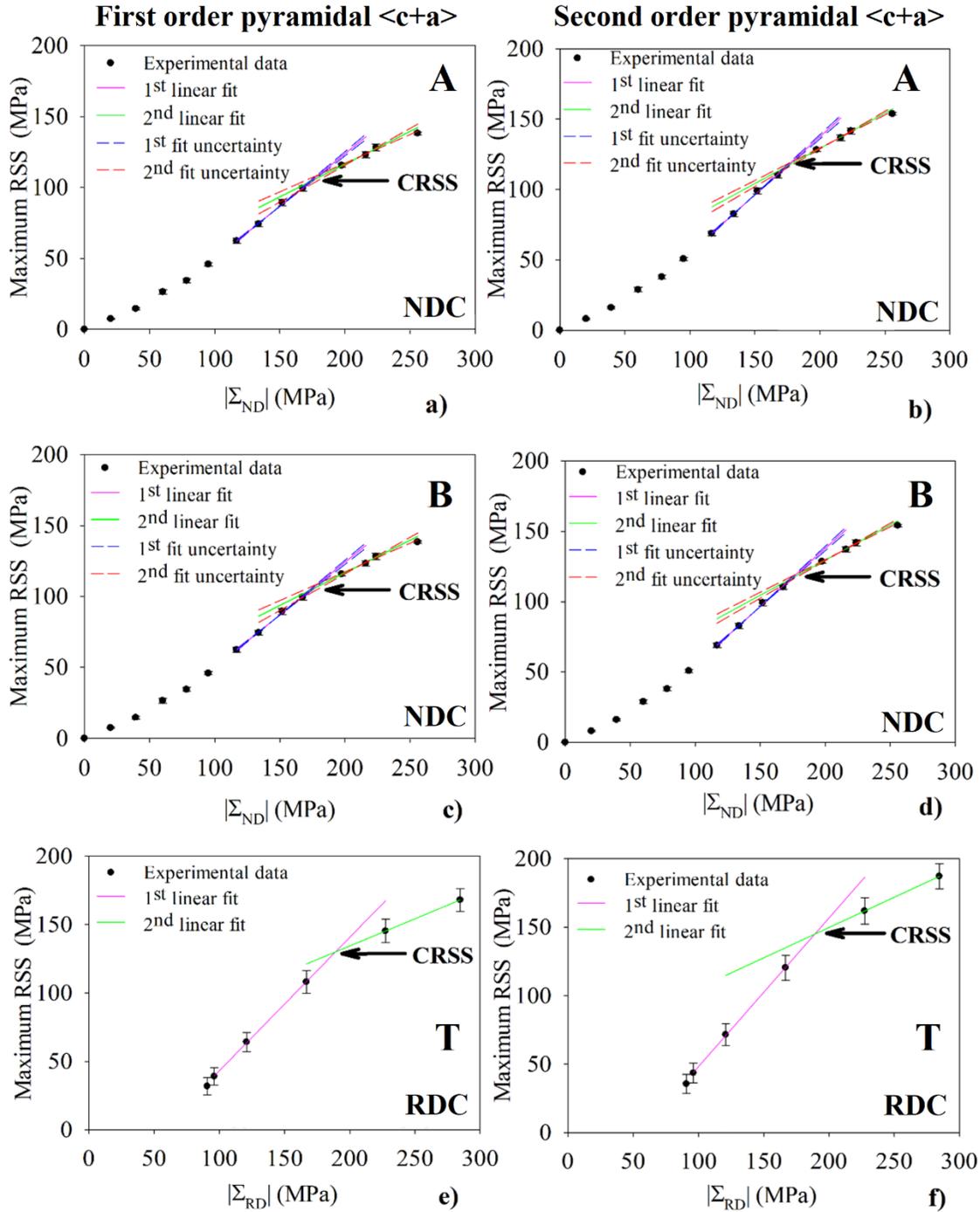

**Fig. 13.** Evolution of the maximum RSS versus macroscopic compression stress shown for pyramidal first order <c+a> systems (on the left) and for pyramidal second order <c+a> systems (on the right). The results from the NDC experiment are shown for A orientation (a, b) and B orientation (c, d). Corresponding results, but for T orientation (twins), obtained from the RDC experiment, are presented (e, f). The error bars correspond to standard uncertainties of RSS (cf. Appendix 2b). The intersection points identified as CRSS values are indicated by arrows.

The values of the CRSS for the pyramidal system <a> and prismatic system can be determined by analysing the behaviour of the RSS in the grains A and B during the tensile test in the RD. In the previous work (cf. [18]), it was found that from the RDT experiment it is possible to estimate the CRSS for the basal system; however, in the case of non-basal systems it could be concluded that they are activated for the RSS in the range of 62-85 MPa. It was not possible to determine which system among the pyramidal <a>, prismatic and pyramidal <c+a> systems is responsible for the initiation of plastic deformation for A- and B-oriented grains, because they all show non-zero values of the RSS for the load applied along the RD. The NDC experiment then demonstrated that the determined CRSS values for pyramidal <c+a> systems are significantly higher than the RSS value on these systems during RDT testing (compare Table 3 and Fig. 13). Therefore, pyramidal <c+a> systems cannot be activated, and activation of pyramidal <a> and/or prismatic systems must be responsible for changing the trend of the RSS versus $E_{RD}$ dependence during the RDT test. The beginning of the change in the curve trend enabled us to determine the CRSS value, which is equal to the RSS at this point. The uncertainty of the CRSS is determined on the basis of the measurement uncertainties of the RSS values that are closest to the trend change. The so determined CRSS values for a pyramidal <a> system and prismatic system are given in Table 3 and shown in Fig. 14. As expected, these values are smaller when compared to CRSS values for the pyramidal <c+a> systems. For this reason, the macroscopic stress-strain curve is significantly lower in the RDT test compared to the NDC test. It is impossible to distinguish which of the systems or both systems, pyramidal <a> and prismatic, are activated, but it is worth noting that both systems have the same slip direction <a>.

Finally, the CRSS for the twin system can be determined from the RSS evolution occurring during compression along the RD. In this case, the orientation B is considered because the Schmid factor for the twin system in this orientation is the highest among all other orientations considered. As shown in Fig. 15, the maximum RSS for the twin system increases linearly and saturates suddenly until the plot finally ends as the B-grains disappear. The CRSS value for the tensile twin system can be easily determined as the arithmetic mean of three RSS for the horizontal part of the graph shown in Fig. 15. The CRSS uncertainty is estimated as the maximum deviation of the RSS from the mean value, extended by its standard uncertainty.

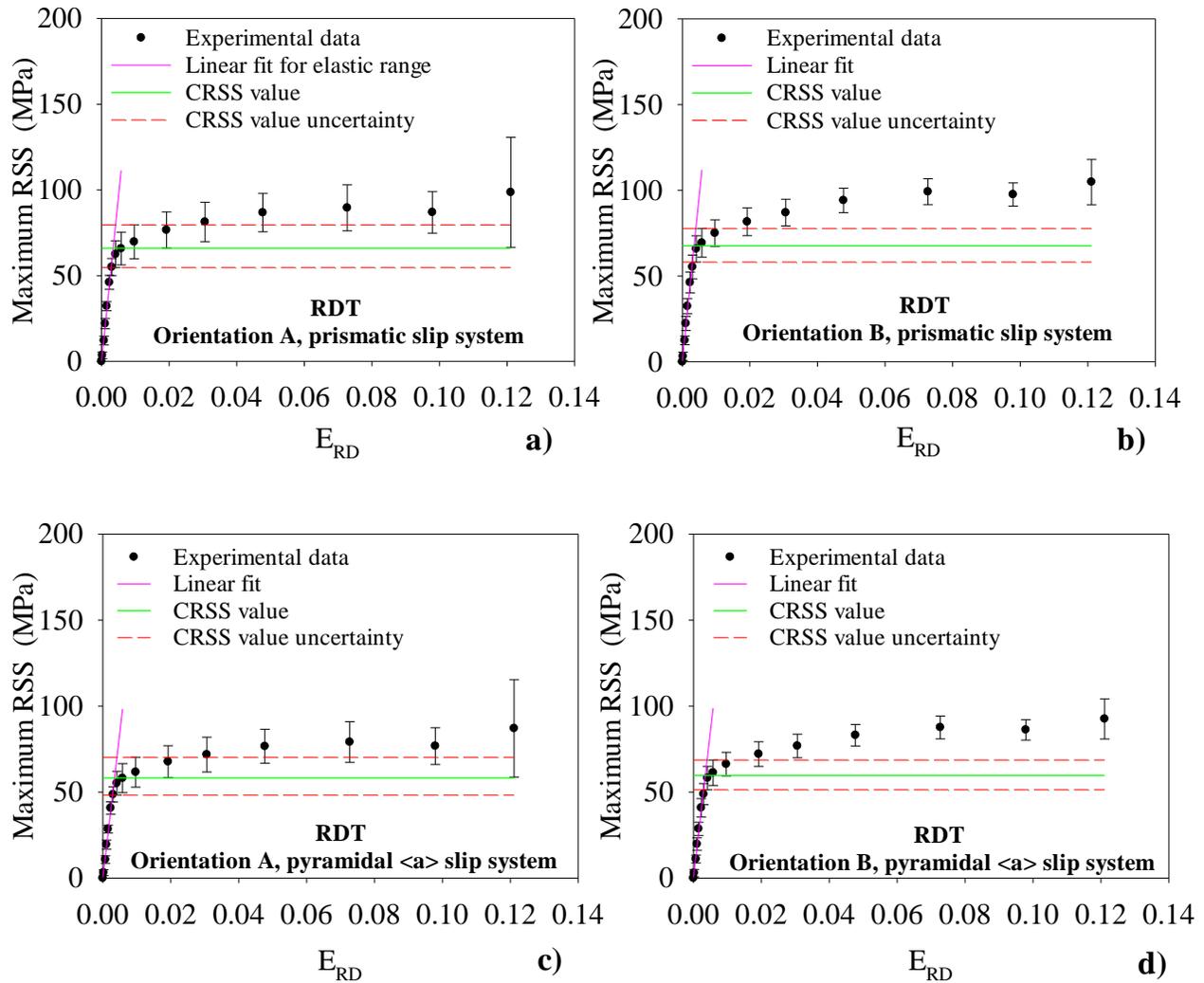

**Fig. 14.** Evolution of the RSS for a prismatic system and pyramidal <a> system (activated in grains having orientations A and B) versus sample strain $E_{RD}$ during the tensile test performed along the RD. The error bars correspond to standard uncertainties of RSS (cf. Appendix 2b). Determined CRSS values and their uncertainties are marked with horizontal lines.

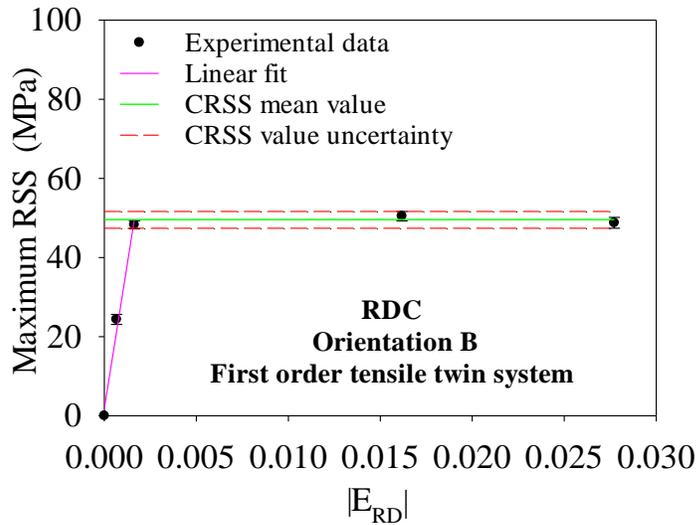

**Fig. 15.** Evolution of the RSS for a tensile twin system (activated in grains having orientations B) versus sample strain $|E_{RD}|$ during the compression test performed along the RD. The error bars correspond to standard uncertainties of RSS (cf. Appendix 2b). Determined CRSS values and their uncertainties are marked with horizontal lines.

The values of the CRSS determined in this work are given in Table 3 and compared with those obtained in [18]. As shown in this table, the previously determined CRSSs are close to the values obtained in this work excluding the values obtained for pyramidal <c+a> systems. Previously, it was found that the CRSS for all systems, excluding the basal system, are between 62-85 MPa, while in this work it was shown that the pyramidal <c+a> systems exhibit a CRSS higher than 100 Mpa. In the new analysis, considering compression tests in the ND and RD, the CRSS for all slip systems and twin systems were determined. However, as mentioned above it is not possible to find out which system, pyramidal <a> or prismatic or both together, are active, and the same concerns the second order pyramidal <c+a> systems.

**Table 3.** CRSS values and their uncertainties determined from experiment and parameters of Voce law used in the modified EPSC model.

| Slip system | | Publication [18] CRSS from experiment - $\tau_0$ (Mpa) | Present work CRSS from experiment – $\tau_0$ (Mpa) | Parameters of Voce law used in modified EPSC model (Mpa) |
|---|---|---|---|---|
| Basal B: $\{0001\}\langle11\bar{2}0\rangle$ (initial sample) | | 35 | 28.0 (3.1) | $\tau_0=28$, $\tau_1\rightarrow0$ $\theta_0\rightarrow0$, $\theta_1\rightarrow0$ |
| Prismatic P: $\{1\bar{1}00\}\langle11\bar{2}0\rangle$ (initial sample) | | 62-85 | 67.7 (7.9) | $\tau_0=68$, $\tau_1=20$ $\theta_0=180$, $\theta_1=50$ |
| Pyramidal <a>: $\{1\bar{1}01\}\langle11\bar{2}0\rangle$ (initial sample) | | | 59.7 (6.9) | $\tau_0=60$, $\tau_1=20$ $\theta_0=180$, $\theta_1=50$ |
| First order pyramidal <c+a> : $\{1\bar{1}01\}\langle\bar{1}2\bar{1}3\rangle$ (initial sample) | | | 104.4 (5.6) | $\tau_0=104$, $\tau_1=110$ $\theta_0=1000$, $\theta_1=80$ |
| Second order pyramidal <c+a> : $\{1\bar{2}12\}\langle\bar{1}2\bar{1}3\rangle$ (initial sample) | | | 116.6 (3.5) | $\tau_0=114$, $\tau_1=110$ $\theta_0=1000$, $\theta_1=80$ |
| First order tensile twin: $\{1\bar{1}02\}\langle\bar{1}101\rangle$ (initial sample) | Continuous approach | - | 49.1 (2.5) | $\tau_0=45$ $\theta_0=650$ |
| | Threshold approach $w^{g,twin}=23\%$ | | | $\tau_0=47$ $\theta_0=650$ |
| First order pyramidal <c+a>: $\{1\bar{1}01\}\langle\bar{1}2\bar{1}3\rangle$ (in the twin) | | - | 130 (between 108 and 145 ) | $\tau_0=125$, $\tau_1=110$ $\theta_0=1000$, $\theta_1=80$ |
| Second order pyramidal <c+a>: $\{1\bar{2}12\}\langle\bar{1}2\bar{1}3\rangle$ (in the twin) | | - | 144 (between 120 and 161) | $\tau_0=135$, $\tau_1=110$ $\theta_0=1000$, $\theta_1=80$ |

## 4. Model calculations compared with experiment

In this work, a modified EPSC model based on the idea of [79] was applied to verify the experimental CRSS ($\tau_0^{g,s}$ – initial) values and to determine the hardening parameters used in the Voce law, approximating CRSS ($\tau_{cr}^{g,s}$ in deformed sample) evolution on system $s$ in grain $g$ [63]:

$$\tau_{cr}^{g,s} = \tau_0^{g,s} + (\tau_1^s + \theta_1^s \Gamma^g)\left(1 - e^{-\frac{\theta_0^s}{\tau_1^s}\Gamma^g}\right), \qquad (2)$$

where: $\Gamma^g$ is the total shear strain on all active slip systems, $\tau_0^s$ is the current value of initial critical resolved shear stress (CRSS) on slip/twin system $s$ in grain $g$, while $\tau_1^s$, $\theta_0^s$ and $\theta_1^s$ are phenomenological hardening parameters for the given slip/twin system $s$ (equal for all grains) which must be found out from available experimental results and/or a model.

Self-consistent calculations [71,79,80] were performed for an input file consisting of 10,000 spherical grains with lattice orientations statistically distributed according to the measured ODF (Fig. 4). The elastic properties of each grain were defined by the single crystal elastic constants given in Section 3.1. For twinning, additional grains were added to the calculations taking into account the six possible variants of tensile twins (i.e., up to six twin variants can potentially be created from the same parent grain). The activation and operation of slip and twin systems were equivalent and based on the Schmid criterion and Voce law (Eq. 2).

The initial CRSS ($\tau_0^{g,s}$) for the slip systems and tensile twin system were determined directly from experiments, as described in the previous sections. However, other parameters of the Voce law describing hardening of slip and twin systems (Eq. 2) cannot be found out without the help of a crystallographic model. Therefore, the experimental $\tau_0^{g,s}$ values (or values close to them within the uncertainties) were used as input data for the EPSC calculations, while the hardening parameters of the Voce law were adjusted to fit the model lattice strains to the experimental ones (measured in the direction of the applied load) and macroscopic stress-strain curves, simultaneously.

It should be emphasised that the same set of parameters of the Voce law was applied for the three tests analysed in this work (Table 3). It was found that to fit the model values to the experimental ones, the interaction tensor $T^{gg}$, describing the interaction of a grain $g$ with effective matrix (for details see [79,81]), should be multiplied by the factor $\alpha = 1.2$, i.e. :

$$T_{ijkl}^{gg,\alpha} = \alpha\, T_{ijkl}^{gg} \qquad (3)$$

where $T_{ijkl}^{gg,\alpha}$ is the modified interaction tensor used to calculate the strain rate localization tensor $A_{ijkl}$ and the $T_{ijkl}^{gg}$ tensor is calculated for the Eshelby spherical inclusion embedded into a homogeneous medium characterised by the macroscopic tangent modulus tensor $L_{ijkl}$.

This means that the differences in the strains between grains are enlarged in comparison to those calculated for the spherical Eshelby inclusion by factor $\alpha$, i.e., in this assumption the modified strain localization tensor $\boldsymbol{A^g}$ is given by:

$$\boldsymbol{A^g} = (\boldsymbol{I} - \alpha\, \boldsymbol{T^{gg}}\, \boldsymbol{\Delta l^g})^{-1} \approx (\boldsymbol{I} + \alpha\, \boldsymbol{T^{gg}}\, \boldsymbol{\Delta l^g}) \qquad (4)$$

where $\boldsymbol{I}$ is a fourth rank identity tensor, and $\boldsymbol{\Delta l^g} = \boldsymbol{l^g} - \boldsymbol{L}$ is the difference between the tangent modulus tensors computed for the grain ($\boldsymbol{l^g}$) and the matrix effective modulus ($\boldsymbol{L}$) [79].

Therefore, the interactions between grains in the modified model are shifted from those obtained by the self-consistent model towards the Sachs approach (assumption of homogeneous stresses and free strains). It should be emphasised that the above assumption is necessary to modify the properties of grain interactions to fit the model to the experiment because the original EPSC model does not correctly predict the processes occurring during elastoplastic deformation of the studied textured Mg alloy.

The twinning phenomena can be included into the model in various ways. In this work, twinning is described using two different approaches:

- "Continuous approximation" means that when the RSS reaches the CRSS value, a twin grain is formed with the same stress state and CRSS as that of its parent grain, but with a new crystal lattice orientation. The twin grain then grows with increasing shear strain on the twin system, thus reducing the parent grain volume.
- In the case of "threshold approximation", when the RSS reaches the CRSS value, the twin system is activated, but no twin is formed yet. The parent grain is subjected to deformation in the same way as in the case of crystallographic slips, i.e., the state of stresses of the parent grain and CRSS on its slip systems change. Twin grains are formed, and their orientations are changed with respect to the parent after the total volume fraction of all twins $w^{g,twin} = 23\%$.

- In both approaches, the volume fraction of twins is proportional to the total shear strain $\Gamma^{g,twin}$ on all twin systems in grain $g$, i.e., twin fraction $w^{g,twin} = \Gamma^{g,twin}/0.13$, [61,82]; including six possible variants of twins created in the same parent grain. However, in the case of an RDC test and a strong basal texture, up to two twin variants are created.

It should be emphasised that the only difference between "continuous" and "threshold" approximations is that in the latter one the twinning is postponed and after the offset of $w^{g,twin} = 23\%$, the same procedure is applied.

The assumption introduced for the parent grains is that the grain stress does not increase when the twin system is active. This means that during twinning, which is approximated by a crystallographic slip on the twin system, the parent grain is perfectly plastic. The self-hardening for the twin system is equal to zero, while the twin slip causes the linear hardening of other systems in the parent grain (hardening is described by one parameter of the Voce law, i.e., $\theta_0$ given in Table 3).

### 4.1 Determination of hardening parameters

The comparison of the model prediction with the experimental results obtained for three different tests are shown in Figs. 16 -18. It is well seen that both the elastic lattice strains measured in the direction of applied load and the macroscopic stress-strain curves are well predicted by the modified EPSC model with Voce parameters given in Table 3. In Figs. 16 -18 two types of macroscopic stress-strain curves are shown: the upper line corresponds to the stress during increasing sample load and the lower line shows the macrostress after relaxation at constant strain. It should be emphasised that the results of the diffraction measurement of lattice deformations correspond to the sample load corresponding to the lower curve. Also, the model of elastic-plastic deformation should be adjusted to the lower curve.

Some disagreement between the predicted and experimental lattice strains is observed in the case of the tensile test in the RD (Fig.16), i.e., qualitative agreement occurs, but the difference between lattice strains measured using different $hkl$ reflections is larger than the model values. This means that the shift to the Sachs model is possibly greater than it was assumed in the modified model with $\alpha = 1.2$ (this was also suggested in the paper [18]). In the case of the compression test in the ND (Fig. 17), a very good agreement between experimental and predicted lattice strains as well as for the macroscopic stress-strain curve was found.

The above-described approximations (continuous and threshold) were used to predict the twinning process during the compression test in the RD. For both approaches, the model results were well adjusted to the experimental results with the same work hardening parameter $\theta_0$ and slightly different CRSS values ($\tau_0$), which are 1-3 MPa lower than the experimental one. In Fig. 18, a comparison of measured lattice strains and macroscopic curve is compared with a model prediction assuming the threshold approach (the results of the continuous approach were very similar).

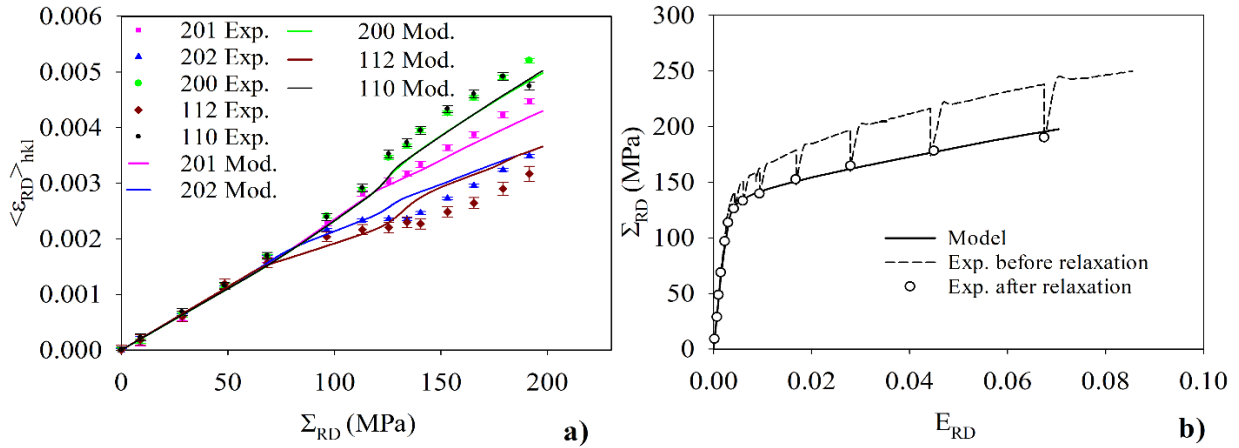

**Fig. 16.** Lattice strains measured in the direction of applied load (a) and macroscopic stress-strain plot (b) compared with the model calculation (threshold approximation) for the tensile test in the RD. The error bars correspond to standard uncertainties of measured lattice strains.

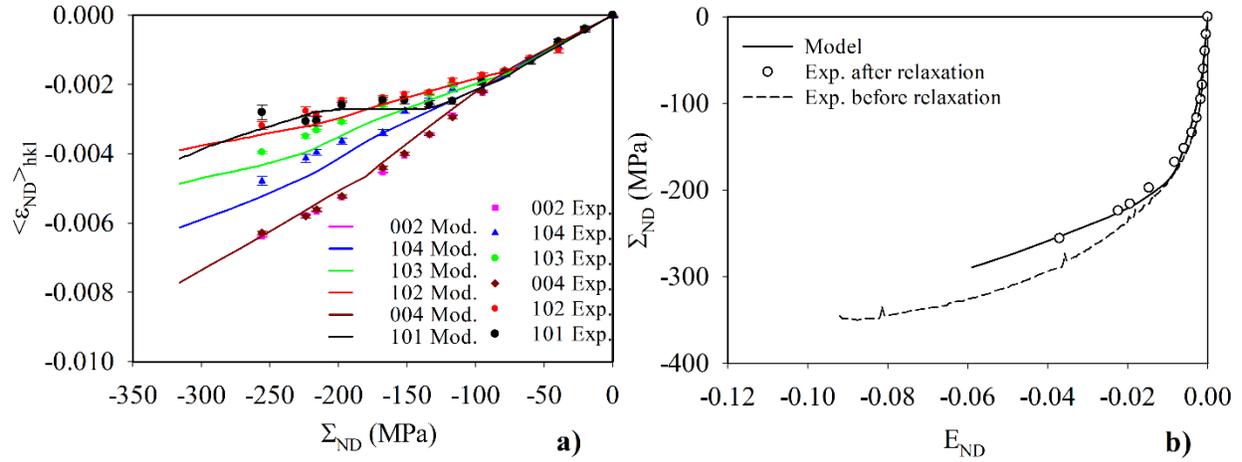

**Fig. 17.** Lattice strains measured in the direction of applied load (a) and macroscopic stress-strain plot (b) compared with the model calculation (threshold approximation) for the compression test in the ND. The error bars correspond to standard uncertainties of measured lattice strains.

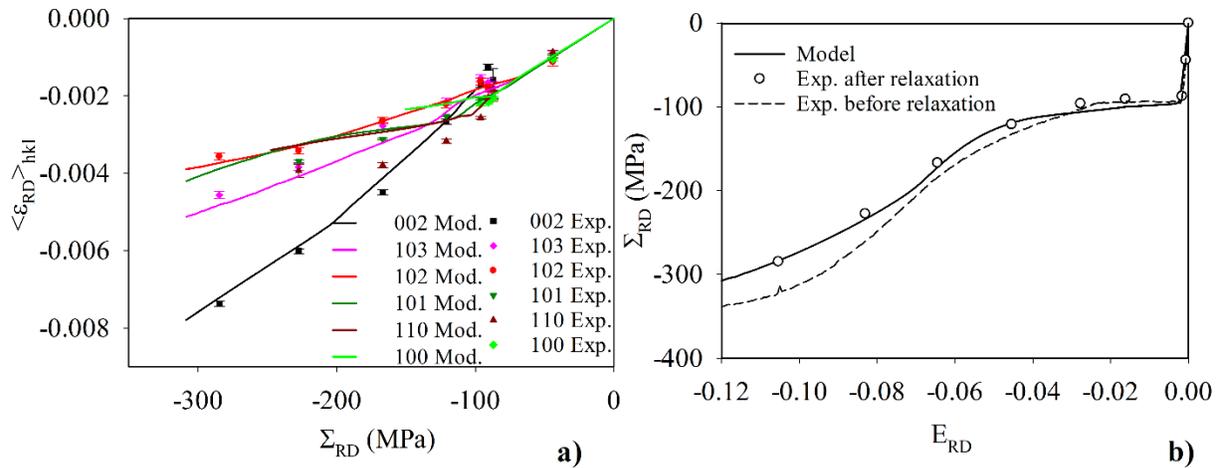

**Fig. 18.** Lattice strains measured in the direction of the applied load (a) and macroscopic stress-strain plot (b) compared with the model calculation for the compression test in the RD with threshold approximation. The error bars correspond to standard uncertainties of measured lattice strains.

### 4.2 Grain stresses and load partitioning

When the model parameters are established, the stress evolution for different grains can be calculated and compared with the experimental results. Such a comparison is shown in Figs. 9-11 (in Section 3.1), where the threshold approximation was used in twinning prediction. It was found that the choice of the type of approximation for twinning prediction is insignificant in the case of the tensile test in the RD and compression test in the ND, because the volume fraction of the twins is negligible. In the case of the compression test in the RD, a slightly better agreement between the model and the experimental stress localized at the twin was found for the threshold approximation.

In general, a very good agreement between experimental and model data was found for the three performed tests and different grain orientations. A significant disagreement was found only for the stress $\sigma_{33}^{C}$ (C orientation) during the tensile test in the RD and for $\sigma_{11}^{A}$ (A orientation) during the compression test in the RD. It should be emphasised that in the above comparison, the experimental data were determined for grain groups (using CGM), while the model results were calculated for a given (single) crystal lattice orientation. Thus, the agreement between the experiment and the model prediction, obtained in most cases, means that the stress tensors determined by the CGM method represent well the grain stresses for given lattice orientations.

To study the partitioning of the load between grains, the evolution of the grain stresses in the direction of applied load together with the macroscopic stress are shown in Figs. 19-21 (the absolute values of stresses and of sample strain are shown). The evolutions of these stresses as a function of the macroscopic stress $|\Sigma_{ND}|$ (or $|\Sigma_{RD}|$) and strain $|E_{ND}|$ (or $|E_{RD}|$) are presented.

As shown in Fig. 19a, during the tensile test along the RD, the grains having orientations A and B (termed "intermediate grains") experience higher stress than the grains with orientations C and D ("soft grains"). This occurs because the basal system, with the lowest value of the CRSS, can only be activated for soft grains where the <c> axis is deviated from the ND direction. The yield stress for the intermediate grains depends on the CRSS of the prismatic and/or pyramidal <a> systems, which are higher than the CRSS for the basal system, but lower than for the pyramidal <c+a> system. It was found that the value of macroscopic stress is between the stresses localized at intermediate and soft grains. The model macrostress value is equal to the mean calculated by the modified EPSC model for all grains included in the model sample (crystallographic texture is taken into account in the orientation distribution of the grains). As shown in Fig. 19b, the macroscopic curve obtained from the model is well fitted to the experimental data, and the qualitative agreement between the experiment and the model was obtained for intermediate grains (arithmetic mean for A and B orientations), as well as soft grains (the arithmetic mean for the orientations C and D). As mentioned above, the quantitative discrepancy indicates that the difference between the loading of hard and soft grains is higher in the real sample compared to the calculations of the modified EPSC model.

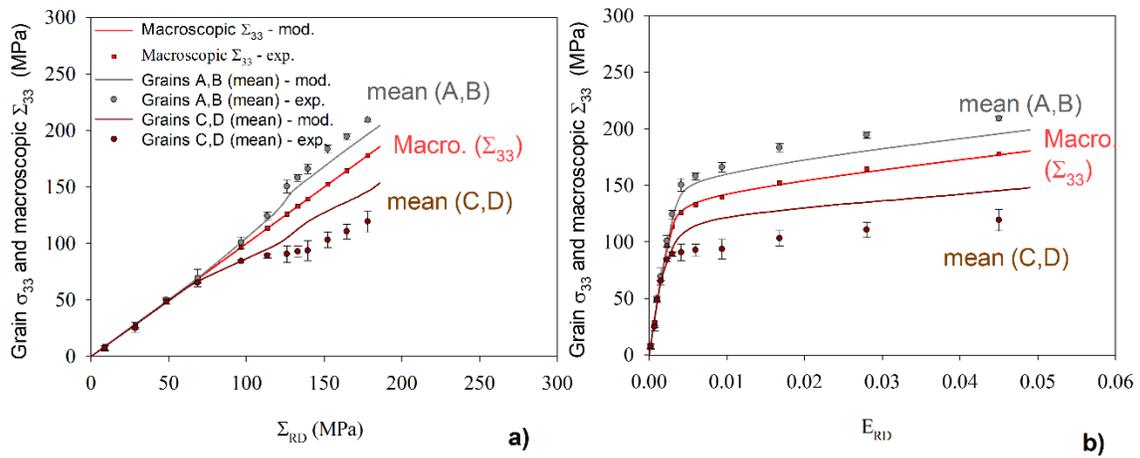

**Fig. 19.** Partitioning of the macroscopic stress $\Sigma_{33}$ between soft (mean for C and D) and intermediate (mean for A and B) grains during the tensile test in the RD. Model predictions are compared with experimental results. The error bars correspond to standard uncertainties of measured lattice strains.

The load partitioning between the hard (A and B) and soft grains (D, F, G) during compression along the ND is presented in Fig. 20a. In this case, a very good agreement between the model and the experiment was obtained. The A and B orientations are now called hard because they carry much more load compared to the moderate A and B orientations in the RDT experiment (cf. Fig. 20 and Fig. 19). In the compression test in the ND, the difference between the yield stress for soft and hard grains is much more significant compared to the tensile test in the RD, because plastic deformation in hard grains occurs as a result of slip on pyramidal systems <c+a> (first or second order), for which the CRSS is the highest of all slip systems (Table 3). Therefore, the macroscopic stress for the sample is higher in the case of compression in the ND compared to the tensile in the RD (cf. macroscopic curves in Fig. 19 and Fig. 20). Due to the large difference between the loading of hard and soft grains, the yield points of soft grains (mean for D, F and G orientation) and of hard grains (mean for A and B orientations) are clearly visible in Fig. 20 as changes in the curve trends at $|\Sigma_{ND}| \approx 70$ Mpa, $|E_{ND}| \approx 0.0015$ (yield stress of the soft grains) and at $|\Sigma_{ND}| \approx 170$ Mpa, $|E_{ND}| \approx 0.008$ (yield stress of the hard grains).

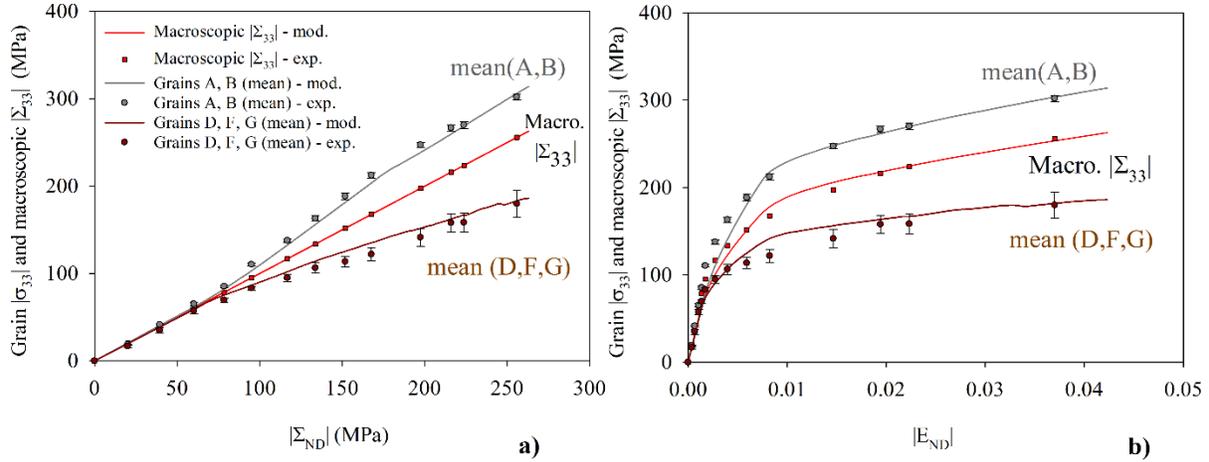

**Fig. 20.** Partitioning of the macroscopic stress $|\Sigma_{33}|$ between soft (mean for D, F and G) and hard (mean for A and B) grains during the compression test in the ND. Model predictions are compared with experimental results. The error bars correspond to standard uncertainties of measured lattice strains.

The load partitioning between grains having different orientations during the compression test in the RD direction is more complex due to the twinning phenomenon. As shown in Fig. 21, all grains are almost equally loaded up to the stress at which the twinning is activated at about $|\Sigma_{RD}| \approx 90$ MPa. The basal system in the soft grain with orientation G' is also activated at about this load. As expected, the stress $|\sigma_{33}^{G'}|$ for this grain is the lowest compared to all other grains. The B-oriented grains transform to twins (T-oriented) and during this transition the macroscopic stress $|\Sigma_{RD}|$ practically remains constant, cf. Fig. 21b, 21d, 21f. This is why in the model, perfect plasticity of parent grains (B and A – oriented) was assumed, which leads to a very small increase of macroscopic stress up to $|E_{RD}| \approx 0.03$ (cf. Fig. 21b, 21d, 21f) as seen in the experimental macroscopic curve. The created twins (T-oriented) exhibit a smaller stress $|\sigma_{33}^{T}|$ compared to other grains, due to back stress generated during twinning (arrow in Fig. 21b, 21d, 21f). It should be however emphasised that the effect of back stress is much smaller than reported in [61]. Because in this version of the model the back stress is not taken into account (the twin grain initially has the same stress state as the parent), the model overestimates the stress for T-oriented grain compared to experimental result at the beginning of twinning process. However, it is worth noting that in the case of threshold approximation the stress localized at the twin grain is lower than in the case of continuous approximation (cf. Fig. 21a-d), because of relaxation of the stresses in the parent grain occurring before the twin is created at the threshold. For the greater load, the model stress at the twin $|\sigma_{33}^T|$ approaches the experimental value, and at the end of the test the agreement between experiment and model is very good and again better accordance was found in the case of the threshold approximation. In the case of A-oriented grains, the stress calculated by the model is lower than the experimental one, but the tendencies of the experimental and model plots agree qualitatively. These grains also transform into twins but stay longer in the sample, i.e., over $|\Sigma_{RD}| = 200$ Mpa.

It should be emphasised that an interesting interplay between the grain stresses occurs, giving the macroscopic sample stress which is an average over all grain stresses. The macroscopic plot obtained by the model is perfectly fitted to the experimental points, and the stress distribution between grains of different orientations is also well predicted by the model (at last qualitatively in the case of orientation A).

In Fig. 21b, 21d, 21e, 21f, the grain stresses $|\sigma_{33}^B|$ and $|\sigma_{33}^T|$ are shown together with the macroscopic stress. It is worth noting that the $|\sigma_{33}^B|$ stress remains constant over the plateau range, i.e., no change in intergranular stress appears in the parent grains during the twinning process and the localized stress remains constant. Due to the lack of initial T-oriented grains in the undeformed sample, the $<d>_{hkl}^0$ value was estimated by interpolation as shown in Supplement 2a (Supplementary materials). It is worth noting that the new T-oriented grains (created due to the twinning process) significantly affect the macroscopic behaviour of the sample. As shown in Fig. 21e, 21f, the shape of the plot $|\Sigma_{RD}|$ versus $|E_{RD}|$ is enforced by the $|\sigma_{33}^T|$ stress evolution, and high macrostress value at the end of the test occurs due to the high stress localized at twins. Thus, the twins exhibit high yield stress and can be considered as the reinforcing grains for the studied sample. In the case of grains with T orientation, uniaxial stress $|\Sigma_{RD}|$ is perpendicular to the base plane (the basal system is not active) and perpendicular to the <a> axis (prismatic and pyramidal <a> are inactive), i.e., only pyramidal <c+a> systems can be activated. The CRSS for the latter system is the highest of all systems (see Table 3).

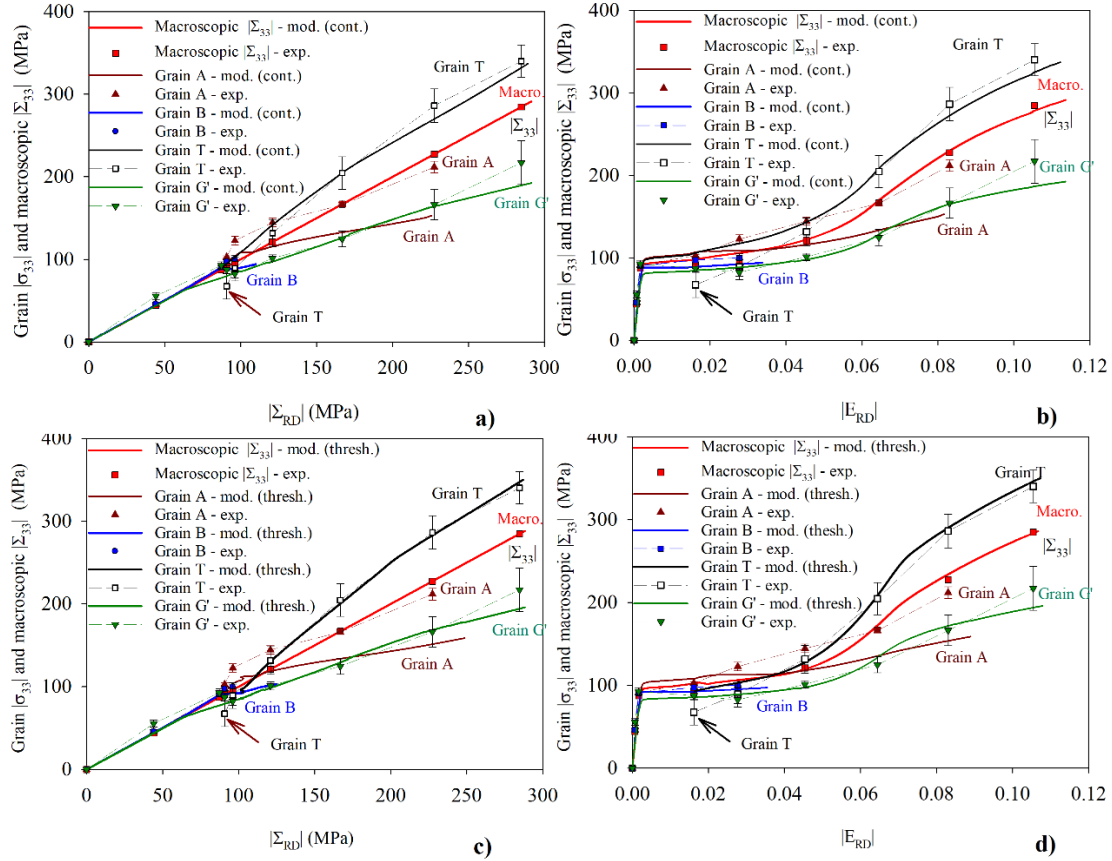

**Fig. 21.** Partitioning of the macroscopic stress $|\Sigma_{33}|$ between soft (mean for C and D), intermediate (A) grains and twins (T) during the compression test in the RD. Two versions of the model prediction, with continuous approximation (a, b) and with threshold approximation (c, d) are compared with experimental results. The error bars correspond to standard uncertainties of measured lattice strains.

### 4.3 Texture evolution

The mechanical behaviour of grains was described in the previous sections. However, it should be emphasised that the macroscopic properties of the sample depend not only on the properties of grains but also on the interaction between them as well as on the number of grains having specific orientations (characterised by texture). The advantage of the diffraction method is the possibility of measurement of texture evolution due to twinning and slips on crystallographic planes. Therefore, in this section the changes in texture (ODF) and volume fraction of twin grains, measured using the EBSD method and neutron diffraction, are compared with the modified EPSC model.

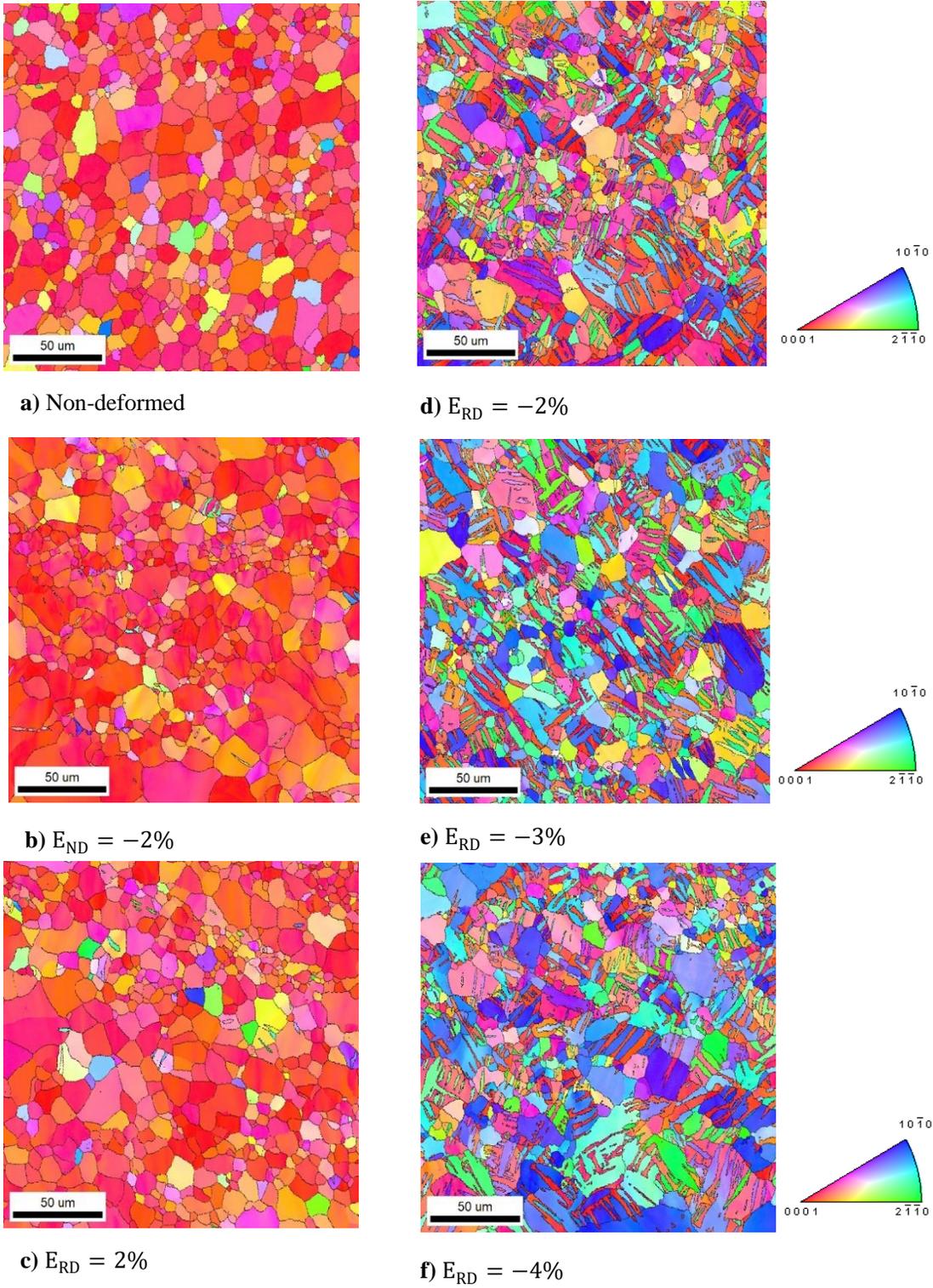

**Fig. 28.** ND - inverse pole figure maps (EBSD) for: a) initial sample, b) compressed to -2% in the normal direction ND, c) stretched to 2% in the rolling direction RD, d) compressed to -2% in the RD, e) compressed up to -3% in the RD, and f) compressed to -4% in the RD.

The ODFs were determined from EBSD measurements for the samples subjected to different tests (Fig. 28). The ODF for the initial undeformed sample is shown in Fig. 4 and the grain orientations that are tested in this work are marked. The changes of texture are considered in model calculations, and they are compared with the experimental ones. In Fig. 29, the model and experimental ODFs after the tensile test in the RD up to $E_{RD} = 3\%$ are shown. It is observed that both model and experiment show an increase in ODF value for B orientation and a decrease for A orientation. These changes do not significantly influence the macroscopic properties of the sample nor the properties of grains because insignificant difference was found in the behaviour of A- and B-oriented grains (similar grain stresses were measured for A and B orientations during the tensile test in the RD, cf. Fig. 9).

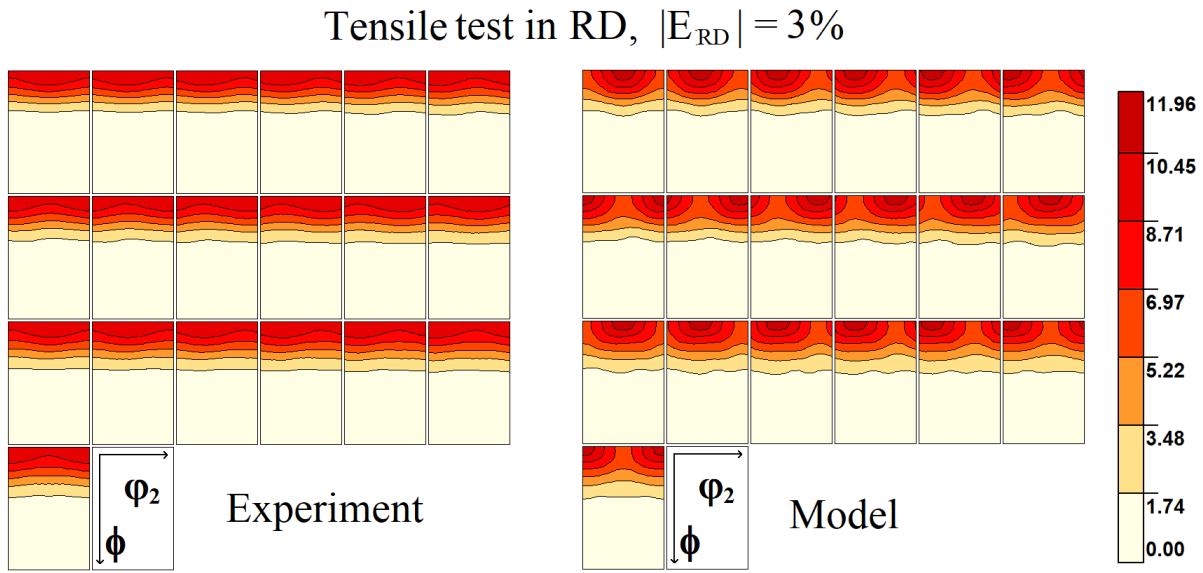

**Fig. 29.** Model predicted and experimental texture after the tensile test in the RD up to $E_{RD}=3\%$. The modified EPSC model with either continuous or threshold approximation give very similar results.

In the case of the compression test in the ND, the changes in the ODF are not significant after deformations up to $E_{ND} = -4\%$ (cf. Fig. 30). Both experiment and model show only small rearrangement of the preferred orientations which does not significantly change the mechanical properties of the sample and the individual grains.

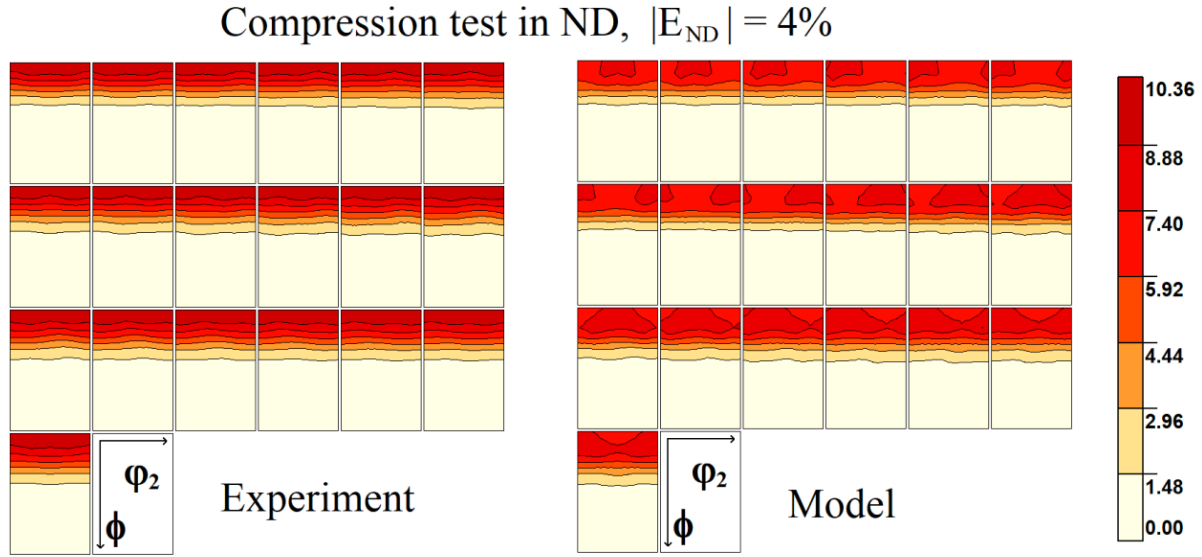

**Fig. 30.** Model predicted and experimental texture after the compression test in the ND up to $E_{ND}$=-4%. The modified EPSC model with either continuous or threshold approximation give practically the same result.

Certainly, the most important changes in ODF occur during the compression test in the RD due to the twinning process. As described above, at first the softening of the sample is caused by twinning, but then the created twins reinforce the sample. The experimental ODF for the deformation $E_{RD} = -4\%$ is compared with the modified EPSC model with continuous and threshold approximations for twin creation in Fig. 31. It was found that the model predicts qualitatively the texture evolution and better agreements between the model and experimental result are obtained for the continuous approximations. It is worth noting that experimental and model results show a significant increase in the twin fraction during the compression test performed in the RD (compare Fig. 31 with Fig. 4, where the twin orientation is marked).

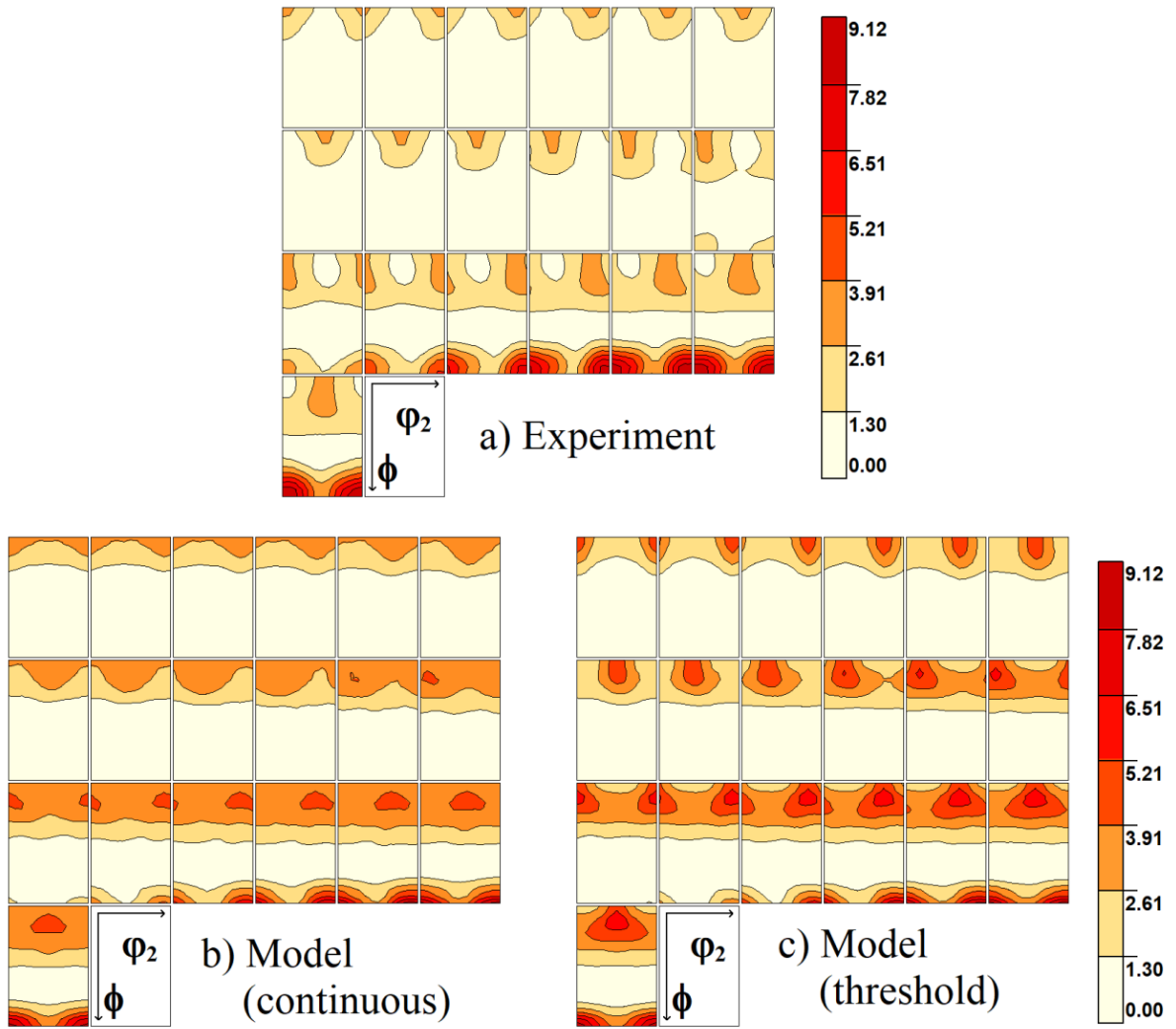

**Fig. 31.** Experimental texture (a) after the compression test in the RD up to $E_{RD}= -4\%$ compared with corresponding results predicted by the EPSC model with continuous (b) and threshold (c) approximation of the twin process.

## 6. Discussion

The main purpose of this work was to develop experimental methods for investigating the micromechanical properties of polycrystalline materials, which enable direct determination of stresses for groups of grains. The neutron diffraction method (CGM) developed in this study showed that the components of the stress tensor can be determined for groups of grains if the lattice strains are determined in many directions. The TOF technique is particularly advantageous because it enables simultaneous determination of lattice deformations with many *hkl* reflections for each direction in which the measurement is performed.

Using neutron diffraction and CGM, the measurements are performed for a large representative volume containing a huge number of polycrystalline grains. It should be emphasised that this method is the only one that can be used for textured polycrystals containing a wide range of grain sizes from a few μm up to hundreds of μm, e.g., a dozen μm as in this work. In contrast, synchrotron diffraction methods with high-energy X-rays can be used to determine stresses for large grains having sizes of at least several tens of μm or larger, which is a significant limitation of these methods. Moreover, the representativity of such measurement is worse in comparison with neutron diffraction, because hundreds of grains having different orientations are available for analysis.

Having developed an experimental method for determination of grain stresses, the measurements can be carried out *in situ* for various loads applied to the samples. These measurements enable experimental studies of the behaviour of polycrystalline grains during elastic-plastic deformation, in particular such phenomena as slips on crystallographic planes, the phenomenon of twinning, stress localization in individual grains, and changes in the crystallographic texture during plastic deformation. In this work, the stresses were determined for groups of grains having different orientations in textured hot-rolled magnesium alloy AZ31. This is a material which, despite its almost isotropic elastic properties, shows a strongly anisotropic response to an applied external load caused by its significant crystallographic texture. Diffraction measurements of lattice strains were carried out *in situ* during the compression/tension of the sample and the stresses for selected groups of crystallites were determined using the CGM.

As mentioned in the Introduction, methodologies for determination of CRSS values from nanoindentation require the use of a crystallographic model [26]. Additionally, the methods used so far, based on measurements of lattice strains, require comparison with models (excluding synchrotron measurements performed for large grains), which makes it impossible to estimate the measurement uncertainty [7,16,22,55,62,65,66,83-90]. One of the most important achievements of the present work is the determination of the values of resolved shear stresses (RSS) and critical shear stresses (CRSS) for various slip systems and twinning directly from the experiment. For this purpose, no model assumptions were used, and the CRSS values were determined from the analysed trends in the evolution of the experimental RSS as a function of macrostress and/or macrostrain values. The values determined from the experiment were obtained with their uncertainties, excluding the CRSS for the <c+a> pyramidal systems for the twin orientation (in the latter case, due to an insufficient number of measuring points, the uncertainty was estimated only roughly). It should be emphasised that the uncertainties of the determined CRSS values were calculated for the first time. The important advantage of the new method presented in this paper is that the results are unambiguous and do not depend on the assumptions used in the model.

Thanks to the stress evolution measurements for different grain groups, it was possible for the first time to determine experimentally the stress distribution between crystallites having different lattice orientations with respect to the applied load. There are four characteristic groups of grains: hard, intermediate, soft, and those for which the twinning phenomenon occurs. Hard grains are those for which the plastic process takes place much later, by activation of the pyramidal slip systems <c+a> showing the highest CRSS value, while the other slips and twinning cannot be activated due to the nearly zero and negative RSS values. Soft grains are those grains in which basal slip having the lowest CRSS value is activated, while the intermediate grains are those for which the prismatic <a> and / or the pyramidal <a> systems are activated, and other slips and twinning remain inactive. In the fourth group of grains, the twinning phenomena can occur, leading to the saturation of grain stress regardless of the slip systems activity.

It is well known that the partitioning of stresses between the grains plays a key role in the macroscopic response of the sample depending on the combination of the type and direction of the applied load with a sharp crystallographic texture. It should be emphasised that in the present work the partitioning between different groups of grains was for the first time determined qualitatively, directly from diffraction experiment. It was shown that the macroscopic behaviour of the sample is always between the high stresses localised at harder grains and low stresses at softer grains (the macroscopic stress is equal to the average over volume of all grains). Due to the different types (compressive or tensile) and orientations of the applied load, most grains with the preferred texture orientations exhibit very different plastic behaviour leading to the macroscopic anisotropy of the sample. As shown in this work, the greatest macroscopic stresses can be achieved when for most crystallites the applied compressive force is parallel to the c-axis, i.e., when the load is parallel to the ND. Due to such geometrical relationships, the value of the RSS greater than zero for A and B orientations can be achieved only for pyramidal slip systems <c+a>, which can be easily calculated from the Schmid law. Therefore, most grains are hard, and the sample is also hard. It has been shown that grains having orientations A and B exhibit higher stresses than macroscopic stresses, while for inclined grains (orientations C and D) the stresses are lower. Different situations take place during the tensile test in the RD, when for the preferred orientations the twin systems remain inactive (RSS is less than zero), the basal system is also inactive (RSS is equal to zero), and the prismatic <a> and/or pyramidal <a> systems activate for the majority of grains (A and B) at an intermediate level of the applied load (much lower than when the compressive load was parallel to ND). Due to the low value of the stress applied to the sample, the RSS values on the <c+a> slip systems are also too low to activate them. This is an intermediate sample response. As a result, the stresses at A and B orientations are higher than the macroscopic value, while the inclined orientations D, E and F exhibit lower values of stresses. In the third experiment (compression in RD), the most complex sample response was observed, and the shape of the macroscopic stress-strain curve is significantly affected by the twinning phenomena. For a small value of the applied load, the twinning process occurred for the preferred texture orientations, leading to saturation of the macroscopic stress-strain curve, i.e., plastic deformation took place without visible hardening. However, as most of the grains jumped to the twin orientations, significant sample hardening occurred as the twin orientation exhibited the c-axis parallel to the tensile load applied along the RD. This means that the orientations of the twin are hard, and the twins can carry the greatest load (higher stress) compared to other grains and macroscopic stress, which has been shown experimentally. The stresses at the T-oriented grains (twins) are better predicted by the modified EPSC model with threshold approximation compared to continuous approximation. In all tests carried out, the lowest stress is always found at inclined soft grains in which the basal system is activated. These grains soften the sample and reduce macrostresses.

To check the agreement of the EPSC model with the obtained experimental results, the measured CRSS values were entered into the calculations as input data. After adjusting the model to the experimental data by changing hardening parameters (used in the Voce law), a full qualitative agreement was achieved in terms of the evolution of lattice strains, stresses at particular grains, and macroscopic stresses. It should be emphasised that to achieve a very good agreement between the three experiments and the corresponding model calculations, the assumption of the self-consistent model was slightly modified by shifting the type of intergranular interaction towards the Sachs assumption. It was found that during plastic deformation, the total incompatibility strains are about 20% greater than those predicted by the self-consistent approach, i.e., weaker intergranular interaction takes place comparing to the self-consistent Eshelby type model.

The most difficult problem in the conducted analysis concerns hardening related to the twin systems. Two approaches were considered in this work. In the first one, the transformation of the parent grains into twins occurs continuously from the beginning and the twin grains obtain the same stress state and CRSS as the parent. However, it is well known that the twinning nucleation does not occur continuously but rapidly as a change of a part of the parent grain into a twin having a different orientation [91]. Therefore, in the second approach, the twin system is initially active, but it does not produce a twin, up to the given threshold. In this case, the hardening of the slip systems occurs, and the twin grain created after the threshold is harder. Moreover, the stress relaxes during the creation of the twin i.e., back stress is produced due to the eigenstrain of the twin transformation. So-called threshold approximation shows much better agreement with the experimental results compared to continuous approximation, and explains the significant initial hardness of the twin grains, which play a role in the reinforcement during the subsequent sample deformation. The threshold is physically explained by some energy barrier that must be overcome to form a twin and has also been used in previous works, e.g. [61]. In this model, the twin is suddenly created when its relative calculated volume reaches 23%.

The evolution of the crystallographic texture was also studied and agreement between the model and experimental texture was found. The evolution of texture was found to be better estimated under the continuous assumption compared to the threshold model.

As known from the literature, the mechanism of twinning is not so easy to describe. Therefore, the main problem of this work was the prediction of the twinning phenomena in model calculations. Especially, the criterion of twinning initialisation in the present work (and usually in literature [61]) is based on Schmid law and a single CRSS value over which most of the systems are activated leading to a plateau on the macroscopic stress-strain plot. However, it is well known that twinning is initialized before the beginning of the plateau, as was determined from acoustic emission experiments, e.g., [39]. This process was not observed in the present work and potentially can be seen in diffraction if the tendency of the RSS is carefully studied during elastic and small plastic deformation, before plateau on the stress-strain plot (this experiment is planned). Moreover, twinning can be initiated by such processes as autocatalytic twin nucleation leading to appearance of shear bands [20]. The latter mechanism of twin nucleation cannot be studied directly by diffraction and Eshelby type models which do not consider spatial heterogeneity of the occurring processes.

It is worth noting that the twin nucleation process and the stress values at the newly formed twins were not accurately predicted using the EPSC model. Therefore, other methods (e.g., visco-elasto-plastic self-consistent VEPSC model) for simulation of twinning process should be tested using the experimental results obtained in this work. Special attention should be paid to the study of the twin formation process at the beginning of deformation and before the plateau observed on the stress-strain plot.

## 7. Conclusions

1) A new experimental methodology based on *in situ* neutron diffraction measurements of the stresses at the groups of grains was proposed and successfully applied for an AZ 31 textured magnesium alloy subjected to three modes of elastoplastic deformation.
2) The different mechanical behaviours were determined for different groups of grains which can be classified as hard, intermediate, and soft, depending on the orientation of the slip and twin systems with respect to the grain stresses.
3) For the first time, RSS evolutions and then CRSS values were determined directly from the diffraction experiment for all activated slip and twin systems, and the uncertainties of these values were estimated.

4) The experimentally determined CRSS values were used in the modified EPSC model (towards a Sachs model) reducing the number of adjusted parameters. As a result, a set of hardening parameters (Voce's laws) was found for which the predicted lattice strains and macroscopic stress-strain plots are closest to the experimental ones, simultaneously for all tests.

5) After determining all the parameters characterizing the slip and twin systems, the evolution of the model texture was compared with that obtained directly from the experiment. A good agreement between the model and experimental lattice strains and macroscopic curve was obtained.

## Acknowledgments

This work was financed by a grant from the National Science Centre, Poland (NCN), No. UMO-2021/41/N/ST5/00394.

P. K. has been partly supported by: the EU Project POWR.03.02.00-00-I004/16; European Union Horizon 2020 research and innovation program under Grant Agreement No. 857470 and the European Regional Development Fund via the Foundation for Polish Science International Research Agenda PLUS program Grant No. MAB PLUS/2018/8.

The research project was partly supported by the program "Excellence initiative – research university" for the AGH University of Science and Technology.

The neutron diffraction experiments used in this work were partly performed during the period 2017-2019 at JINR in Dubna (Russia) and the purchase of samples/reagents/ancillary equipment was partly financed by the joint JINR/AGH projects No. PWB/389-24/2017, PWB/254_24/2018 and PWB/129-23/2019.

Measurements partly were carried out at the CANAM infrastructure of the NPI CAS Řež. The employment of the CICRR infrastructure supported by MEYS project LM2023041 is acknowledged.

## Data availability

The raw and processed data required to reproduce results presented in this publication are available on https://doi.org/10.58032/AGH/RBP5WX

**Appendix 1.    The set of poles used for stress determination by using the crystallite group method.**

**Table A1.1** Examined poles for crystal orientations A - G' of AZ31 magnesium during normal direction compression (NDC), rolling direction tensile (RDT), and rolling direction compression (RDC) tests.

| No. | (*hkil*) | Ψ [°] | φ [°] | No. | (*hkil*) | Ψ [°] | φ [°] |
|---|---|---|---|---|---|---|---|
| **Orientation A, Experiment: NDC** | | | | **Orientation B, Experiment: RDT** | | | |
| 1 | 0001 | 0.00 | 0.00 | 1 | 10$\bar{1}$0 | 0.00 | 0.00 |
| 2 | $1\bar{2}10$ | 90.00 | 0.00 | 2 | $11\bar{2}0$ | 30.00 | 0.00 |
| 3 | $1\bar{1}02$ | 46.00 | 35.00 | 3 | $01\bar{1}0$ | 60.00 | 0.00 |
| 4 | $10\bar{1}4$ | 14.81 | 97.46 | 4 | $\bar{1}2\bar{1}0$ | 90.00 | 0.00 |
| 5 | $\bar{1}010$ | 90.00 | 90.00 | 5 | $0001$ | 90.00 | 90.00 |
| 6 | $0\bar{1}1\bar{2}$ | 46.00 | 154.08 | 6 | $01\bar{1}3$ | 74.60 | 61.60 |
| 7 | $\bar{1}2\bar{1}0$ | 90.00 | 180.00 | 7 | $10\bar{1}3$ | 58.00 | 90.00 |
| 8 | $1\bar{1}0\bar{2}$ | 46.00 | 205.92 | **Orientation C. Experiment: RDT** | | | |
| 9 | $10\bar{1}\bar{4}$ | 14.81 | 263.00 | 1 | $1\bar{2}12$ | 0.00 | 0.00 |
| 10 | $10\bar{1}0$ | 90.00 | 270.00 | 2 | $01\bar{1}3$ | 88.47 | 74.61 |
| 11 | $0\bar{1}12$ | 46.00 | 334.08 | 3 | $0001$ | 60.00 | 90.00 |
| **Orientation B. Experiment: NDC** | | | | 4 | $20\bar{2}1$ | 82.60 | 13.00 |
| 1 | $0001$ | 0.00 | 0.00 | 5 | $10\bar{1}1$ | 76.36 | 24.77 |
| 2 | $1\bar{2}10$ | 90.00 | 90.00 | **Orientation D. Experiment: RDT** | | | |
| 3 | $1\bar{1}02$ | 46.00 | 125.00 | 1 | $\bar{1}013$ | 90.00 | 90.00 |
| 4 | $10\bar{1}4$ | 14.81 | 187.46 | 2 | $10\bar{1}1$ | 0.00 | 0.00 |
| 5 | $\bar{1}010$ | 90.00 | 180.00 | **Orientation A. Experiment: RDC** | | | |
| 6 | $0\bar{1}1\bar{2}$ | 46.00 | 244.08 | 1 | $1\bar{2}10$ | 0.00 | 0.00 |
| 7 | $\bar{1}2\bar{1}0$ | 90.00 | 270.00 | 2 | $10\bar{1}0$ | 90.00 | 0.00 |
| 8 | $1\bar{1}0\bar{2}$ | 46.00 | 295.92 | 3 | $\bar{1}010$ | 90.00 | 0.00 |
| 9 | $10\bar{1}\bar{4}$ | 14.81 | 353.00 | 4 | $11\bar{2}2$ | 60.00 | 215.00 |
| 10 | $10\bar{1}0$ | 90.00 | 0.00 | 5 | $2\bar{1}\bar{1}2$ | 60.00 | 35.00 |
| 11 | $0\bar{1}12$ | 46.00 | 64.08 | 6 | $0001$ | 90.00 | 90.00 |
| **Orientation D. Experiment: NDC** | | | | 7 | $000\bar{1}$ | 90.00 | 270.00 |
| 1 | $1\bar{1}0\bar{2}$ | 14.81 | 262.53 | **Orientation B. Experiment: RDC** | | | |
| 2 | $1\bar{1}03$ | 0.00 | 0.00 | 1 | $20\bar{2}\bar{1}$ | 14.81 | 97.47 |
| 3 | $10\bar{1}3$ | 60.00 | 125.26 | 2 | $20\bar{2}1$ | 14.81 | 262.53 |
| 4 | $2\bar{1}\bar{1}0$ | 60.00 | 305.26 | 3 | $10\bar{1}0$ | 0.00 | 0.00 |
| 5 | $0\bar{1}13$ | 60.00 | 55.00 | 4 | $1\bar{2}10$ | 90.00 | 0.00 |
| 6 | $1\bar{2}10$ | 60.00 | 235.00 | 5 | $\bar{1}2\bar{1}0$ | 90.00 | 180.00 |
| 7 | $1\bar{1}01$ | 90.00 | 270.00 | 6 | $1\bar{1}02$ | 71.30 | 136.00 |
| 8 | $1\bar{1}01$ | 90.00 | 90.00 | 7 | $1\bar{1}01$ | 60.00 | 215.00 |
| 9 | $11\bar{2}0$ | 90.00 | 0.00 | 8 | $1\bar{1}0\bar{2}$ | 71.30 | 226.00 |
| 10 | $\bar{1}\bar{1}20$ | 90.00 | 180.00 | 9 | $0\bar{1}12$ | 71.30 | 316.00 |

| | Orientation F. Experiment: NDC | | | 10 | 0$\bar{1}1\bar{2}$ | 71.30 | 46.00 |
|---|---|---|---|---|---|---|---|
| 1 | 2$\bar{2}$0$\bar{1}$ | 14.81 | 262.53 | 11 | 0$\bar{1}$11 | 60.00 | 35.00 |
| 2 | 1$\bar{1}$0$\bar{1}$ | 0.00 | 0.00 | 12 | 0001 | 90.00 | 90.00 |
| 3 | 1$\bar{1}$0$\bar{2}$ | 14.81 | 97.47 | 13 | 000$\bar{1}$ | 90.00 | 270.00 |
| 4 | 10$\bar{1}$1 | 47.22 | 25.92 | | Orientation D. Experiment: RDC | | |
| 5 | 0$\bar{1}$1$\bar{1}$ | 47.22 | 154.00 | 1 | 10$\bar{1}$1 | 0.00 | 0.00 |
| 6 | 10$\bar{1}$1 | 71.29 | 315.81 | 2 | 10$\bar{1}$2 | 14.80 | 97.00 |
| 7 | 0$\bar{1}$11 | 71.29 | 225.81 | 3 | 0$\bar{1}$13 | 47.20 | 116.00 |
| 8 | 11$\bar{2}$0 | 90.00 | 0.00 | 4 | 0$\bar{1}$14 | 47.20 | 116.00 |
| 9 | $\bar{1}$1$\bar{2}$0 | 90.00 | 180.00 | 5 | 0$\bar{1}$13 | 60.00 | 125.00 |
| | Orientation G. Experiment: NDC | | | 6 | 10$\bar{1}$3 | 90.00 | 270.00 |
| 1 | 1$\bar{1}$0$\bar{2}$ | 0.00 | 0.00 | 7 | 1$\bar{1}$03 | 47.20 | 63.00 |
| 2 | 1$\bar{1}$0$\bar{1}$ | 14.80 | 277.47 | 8 | 1$\bar{1}$04 | 47.20 | 63.00 |
| 3 | 11$\bar{2}$0 | 90.00 | 0.00 | 9 | 10$\bar{1}$3 | 90.00 | 270.00 |
| 4 | $\bar{1}$1$\bar{2}$0 | 90.00 | 180.00 | 10 | 10$\bar{1}$3 | 90.00 | 90.00 |
| 5 | 10$\bar{1}$2 | 71.29 | 135.81 | | Orientation G'. Experiment: RDC | | |
| 6 | 0$\bar{1}$12 | 71.00 | 44.00 | 1 | 10$\bar{1}$2 | 0.00 | 0.00 |
| 7 | 10$\bar{1}$3 | 60.00 | 125.26 | 2 | 10$\bar{1}$1 | 14.81 | 262.53 |
| 8 | 1$\bar{2}$12 | 88.10 | 255.31 | 3 | 10$\bar{1}$3 | 14.81 | 97.00 |
| | Orientation A. Experiment: RDT | | | 4 | 10$\bar{1}$4 | 14.81 | 97.00 |
| 1 | 1$\bar{2}$10 | 0.00 | 0.00 | 5 | 1$\bar{1}$03 | 60.00 | 125.00 |
| 2 | 1$\bar{1}$00 | 30.00 | 0.00 | 6 | 1$\bar{1}$0$\bar{4}$ | 60.00 | 125.00 |
| 3 | 2$\bar{1}\bar{1}$0 | 60.00 | 0.00 | 7 | 10$\bar{1}\bar{2}$ | 90.00 | 270.00 |
| 4 | 10$\bar{1}$0 | 90.00 | 0.00 | 8 | 0$\bar{1}$13 | 60.00 | 55.00 |
| 5 | 0001 | 90.00 | 90.00 | 9 | 0$\bar{1}$14 | 60.00 | 55.00 |
| 6 | 1$\bar{1}$03 | 62.70 | 72.70 | 10 | 10$\bar{1}\bar{2}$ | 90.00 | 90.00 |
| 7 | 10$\bar{1}$3 | 90.00 | 58.00 | | Orientation F. Experiment: RDC | | |
| 8 | 10$\bar{1}$1 | 90.00 | 28.00 | 1 | 10$\bar{1}$3 | 0.00 | 0.00 |
| | | | | 2 | 10$\bar{1}$2 | 14.81 | 262.53 |
| | | | | 3 | 1$\bar{1}$04 | 47.20 | 116.00 |
| | | | | 4 | 1$\bar{1}$03 | 60.00 | 125.00 |
| | | | | 5 | 1$\bar{2}$10 | 90.00 | 180.00 |
| | | | | 6 | 10$\bar{1}\bar{1}$ | 90.00 | 270.00 |

| | | | |
|---|---|---|---|
| 7 | $\bar{1}2\bar{1}0$ | 90.00 | 0.00 |

## Appendix 2. Strain and stress determination – problem uncertainty analysis

### a) Determination of the interplanar spacings for reflections absent in the initial sample

In the study of AZ31 alloy, the lattice strains were determined based on Equation 1 (in the publication) in which the difference between interplanar spacing in the sample under load ($<d>_{hkl}$) and those in the initial sample were calculated ($<d>_{hkl}^{0}$). This minimises the possible systematic errors in peak position determination. However, in the case of the twin grains arising during compression in the RD, the grains with orientations corresponding to the twins were not present in the initial sample, therefore the necessary interplanar spacings $<d>_{hkl}^{0}$ were not available. This especially concerns the absence of diffraction peaks for 002 and 004 reflections in the detector L2 measuring the interplanar spacings in the direction of the load applied along the RD. Thus, the values of the $<d>_{002}^{0,L2}$ and $<d>_{004}^{0,L2}$ were found using the interpolation method, in which the available interplanar spacings $<d>_{hkl}^{0,L2}$ and $<d>_{hkl}^{0,L5}$ measured respectively by L2 and L5 detectors were used. To do this, the ratios of interplanar spacings for the corresponding reflections $hkl$ were calculated (except for reflections 002 and 004):

$$x_{hkl} = \frac{<d>_{hkl}^{0,L2}}{<d>_{hkl}^{0,L5}}. \qquad (AS2.1)$$

Then, linear regression was used to fit the straight lines, assuming the dependence:

$$x_{hkl} = A <d>_{hkl}^{0,L5} + B \qquad (A2.2)$$

The unknown $x_{002}$ and $x_{004}$ values and their standard uncertainties were found from the fitted lines (Eq. S2.2) for measured $<d>_{002}^{0,L5}$ and $<d>_{004}^{0,L5}$ values as shown in Fig. A2.1. Finally, the values:

$$<d>_{002}^{0,L2} = x_{002} <d>_{002}^{0,L5} \text{ and } <d>_{004}^{0,L2} = x_{004} <d>_{004}^{0,L5} \qquad (A2.3)$$

were found.

The combined standard uncertainty of the determined interplanar spacing $u(<d>_{002}^{0,L2})$ was calculated from standard uncertainties $u(x_{002})$ and $u(<d>_{002}^{0,L5})$, using the law of propagation of uncertainty (LPU). In the same way, the uncertainty $u(<d>_{004}^{0,L2})$ was found. The procedure described above was applied for the data obtained before ($\omega = 0°$) and after sample rotation about the sample axis ($\alpha = 90°$) (Fig. A2.1), and the results are given in Table .

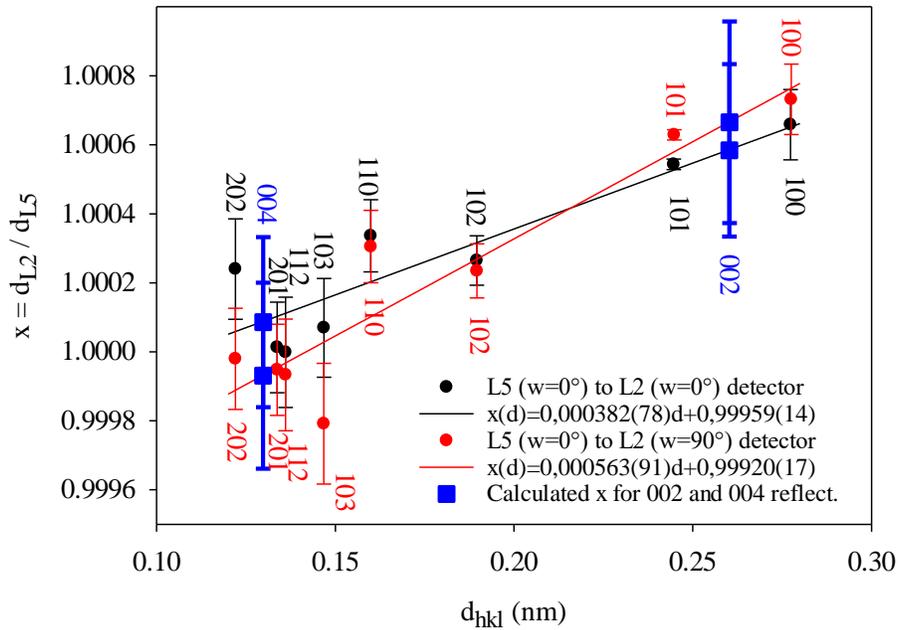

Fig. A2.1 The dependences of $x_{hkl}$ vs. $<d>_{hkl}^{0,L5}$ fitted by straight lines based on which the $x_{002}$ and $x_{004}$ values were interpolated for the measured $<d>_{002}^{0,L5}$ and $<d>_{004}^{0,L5}$ values.

Table S2.1 Determined $<d>_{002}^{0,L2}$ and $<d>_{004}^{0,L2}$ interplanar spacings for the sample before rotation ($\alpha = 0°$) and rotated around the RD$\|x_2$ axis by $\alpha = 90°$.

| Reflection | $\alpha$ (°) | $<d>_{hkl}^{0,L2}$ (nm) | $u(<d>_{hkl}^{0,L2})$ (nm) |
|---|---|---|---|
| 002 | 0° | 0.260232 | 0.000065 |
| 004 | 0° | 0.129791 | 0.000032 |
| 002 | 90° | 0.260253 | 0.000076 |
| 004 | 90° | 0.129771 | 0.000035 |

**b) Stress and CRSS determination - uncertainty analysis**

The procedure is based the normalized goodness of fit (GoF) $\chi^2$, which is defined as:

$$\chi^2 = \frac{1}{N-M}\sum_{n=1}^{N}\left(\frac{\langle\varepsilon(\varphi_n,\psi_n)\rangle_{hkl}^{exp}-\langle\varepsilon(\varphi_n,\psi_n)\rangle_{hkl}^{cal}}{u(\langle\varepsilon(\varphi_n,\psi_n)\rangle_{hkl}^{exp})}\right)^2, \qquad (A2.4)$$

where $\langle\varepsilon(\varphi_n,\psi_n)\rangle_{(hkl)}^{exp}$ and $\langle\varepsilon(\varphi_n,\psi_n)\rangle_{hkl}^{cal}$ are respectively the experimental and calculated lattice parameters, $u(\langle\varepsilon(\varphi_n,\psi_n)\rangle_{hkl}^{exp})$ is the standard uncertainty of $\langle\varepsilon(\varphi_n,\psi_n)\rangle_{hkl}^{exp}$ for the *n-th* measurement (obtained from procedure for analysis of diffraction data), and *N* and *M* are the number of measured lattice strains and number of fitting parameters, respectively.

In the stress analysis performed in this work, the values of lattice strains given by Eq. 1 are fitted to experimental data using the General Linear Least Squares (GLLS) method in which the solution is obtained by applying the Singular Value Decomposition (SVD) procedure described by Press et al. [74]. In this procedure, the stress components $\sigma_{ij}^{CR}$ are considered as model adjustable parameters which are estimated by fitting the model (c.f. Eq. 1) to available experimental lattice strains using GLLS analysis. The standard uncertainties of so obtained stresses $\sigma_{ij}^{CR}$ are calculated as the square roots from the obtained variances $u^2(\sigma_{ij}^{CR})$ multiplied by $\chi^2$, i.e., $u(\sigma_{ij}^{CR}) = \sqrt{u^2(\sigma_{ij}^{CR})*\chi^2}$ [74]. Such an estimation is made to take into account the impact of uncertainties $u(\langle\varepsilon(\varphi_n,\psi_n)\rangle_{hkl}^{exp})$ in the obtained results through the definition of $\chi^2$ given by Eq. A2.4. The so obtained uncertainties are presented as the error bars in Figs. 9-11.

The uncertainties of the RSS values were calculated as the combined standard uncertainties calculated from the $u(\sigma_{ij}^{CR})$ uncertainties and they are presented in Figs. 22-27, Figs. 12-15 and Figs. A3.1-A3.5 as the error bars.